# Experimental topological quantum computing with electric circuits

Deyuan Zou [1*], Naiqiao Pan[*], Tian Chen[1*], Houjun Sun[2], and Xiangdong Zhang[1+]

[1] Key Laboratory of advanced optoelectronic quantum architecture and measurements of Ministry of Education, Beijing Key Laboratory of Nanophotonics & Ultrafine Optoelectronic Systems, School of Physics, Beijing Institute of Technology, 100081, Beijing, China

[2] Beijing Key Laboratory of Millimeter wave and Terahertz Techniques, School of Information and Electronics, Beijing Institute of Technology, Beijing 100081, China

*These authors contributed equally to this work. [+$]Author to whom any correspondence should be addressed. E-mail: zhangxd@bit.edu.cn, chentian@bit.edu.cn

**Abstract: The key obstacle to the realization of a scalable quantum computer is overcoming environmental and control errors[1-3]. Topological quantum computation has attracted great attention because it has emerged as one of the most promising approaches to solving these problems[4-8]. Various theoretical schemes for building topological quantum computation have been proposed[9-24]. However, experimental implementation has always been a great challenge because it has proved to be extremely difficult to create and manipulate topological qubits in real systems. Therefore, topological quantum computation has not been realized in experiments yet. Here, we report the first experimental realization of topological quantum computation with electric circuits. Based on our proposed new scheme with circuits, Majorana-like edge states are not only observed experimentally, but also T junctions are constructed for the braiding process. Furthermore, we demonstrate the feasibility of topological quantum computing through a set of one- and two-qubit unitary operations. Finally, our implementation of Grover's search algorithm demonstrates that topological quantum computation is ideally suited for such tasks.**

Quantum computing has attracted much attention in recent decades, since it has shown great promise as a resource providing exponential speedup over classical computers for certain problems[1-3, 25, 26]. Building a practical quantum computer has always been the goal of people. However, it is notoriously hard to build such a device due to the ubiquitous decoherence of quantum states and error rendering. At present, there are two propositions to solve these problems. One way is to use logical qubits and error-correcting codes to correct all errors that appear during computation[27-35]. Such a scheme is extremely difficult to implement because it requires a lot of additional resources for quantum circuits. Another way is to perform a topological quantum computing scheme.

Topological quantum computation aims to employ anyonic quasiparticles with exotic braiding statistics to encode and manipulate quantum information in a fault-tolerant way[4-8]. Majorana-like zero modes are experimentally the simplest realization of anyons that can non-trivially process quantum information. Based on the scheme of Majorana-like zero modes, some potential realizable systems have been analyzed theoretically, such as superconductor and fractional quantum Hall liquid[9-15], semiconducting heterostructures and wires[16-20], photonic networks[21-23], electric circuits[24] and so on. However, the experimental implementation has encountered great challenges. For example, according to the circuit scheme as described in Ref[24], not only do the variable capacitor and inductor need to be precisely regulated at the same time, but also the whole braiding operation needs to be completed in a very short time. This is because the input signal decays quickly due to the loss of circuit components. Therefore, it is very difficult to realize in experiments. How to realize topological quantum computation experimentally is still an open problem.

In this work, we propose a new circuit scheme to overcome the problems of the previous schemes. We construct a resistor-capacitor (RC) circuit instead of an LC circuit. In our designed RC circuits, the task can be completed only by adjusting the resistance, without precise adjustment of capacitances and inductances at the same time. Based on such circuits, we observe both theoretically and experimentally Majorana-like edge states and construct T junctions for the braiding process. In addition, we use segmented fixed resistances to replace variable resistances, which allows the whole braiding process to be completed before the signal is attenuated. A set of one- and two-qubit unitary operations is realized experimentally by the braiding operations. Furthermore, we fabricate integrated circuits for the topological quantum computation to implement Grover's search algorithm.

**Electric-circuit realization of non-Abelian braiding of Majorana-like edge states**

Constructing Majorana-like edge states and braiding process are the key steps to realize topological quantum computation. Majorana-like edge states can be constructed by various models[6-8, 12-14, 16-24]. The Kitaev model is the fundamental one-dimensional model hosting Majorana-like edge states. The Hamiltonian of the Kitaev model is expressed as

$$H_K(k) = \frac{1}{2}\begin{pmatrix} \varepsilon_k & i\Delta e^{-i\phi'}\sin k \\ -i\Delta e^{i\phi'}\sin k & -\varepsilon_k \end{pmatrix}$$ with $\varepsilon_k = -t\cos k - \mu$, where $t$, $\mu$, $\phi'$ and $\Delta$

represent the hopping amplitude, chemical potential, superconducting phase and gap parameters, respectively. It is well known that the system is topological for $|\mu| < |2t|$ and trivial for $|\mu| > |2t|$. A pair of topological zero-energy states emerges at the edges of a topological phase according to the bulk-edge correspondence. They are protected by particle-hole symmetry. Their properties are identical to Majorana edge states in p-wave superconducting. The theoretical correspondence is shown in Supplementary Note 1.

Now, we discuss how to design the electric circuit to simulate such a model. The designed electric circuit network can be described with the help of Fig. 1(a). Fig. 1(a) shows the electric circuit with two units corresponding to the Kitaev model, each unit cell contains four circuit nodes (A-D). The advantage of the circuit is that we only need to consider the connection relationship between different circuit nodes instead of their location[36-47]. The blue and red lines in Fig. 1(a) represent the connections between nodes in two units, which correspond to two kinds of intercell couplings. The detailed connection modes are described in Fig. 1(b). The blue and red arrows in Fig.1(b) represent negative impedance converters (INIC) through current inversion with different resistances ( $R_a$ and $R_b$ ), where their structures are shown in the blue and red boxes of Fig. 1(c). The green arrows in Fig. 1(a) represent intracell coupling, which the INIC structure with the resistance $R_c$ is shown in the green box of Fig. 1(c). The red and blue dots on the green lines in Fig. 1(a) are switches that can determine whether the coupling exists or not, which are displayed in the black box of Fig. 1(c). Besides, proper grounding elements should be connected to each node. Details of the grounding parts are shown in Supplementary Note 2. For the above circuit network, we can derive a circuit Laplacian by Kirchhoff's law:

$$J = \begin{pmatrix} \operatorname{Im} J_K(k) & -\operatorname{Re} J_K(k) \\ \operatorname{Re} J_K(k) & \operatorname{Im} J_K(k) \end{pmatrix}, \quad (1)$$

where $J_K(k) = \dfrac{1}{2C} \begin{pmatrix} \varphi_k & 4iR_a^{-1}e^{-i\phi}\sin k \\ -4iR_a^{-1}e^{i\phi}\sin k & -\varphi_k \end{pmatrix}$ with $\phi = 0$, $\varphi_k = -4R_b^{-1}\cos k - 2R_c^{-1}$, $C$ is capacitance connecting to the ground. Here $\phi$ is circuit phase which determines the change of variable resistor, corresponding to the 'superconductor phase $\phi'$' in the Kitaev model. If we choose the parameters to be $(R_a C)^{-1} = \dfrac{\Delta}{4}$, $(R_b C)^{-1} = \dfrac{t}{4}$ and $(R_c C)^{-1} = \dfrac{\mu}{2}$. Then, the $J_K(k)$ in the circuit Laplacian $J$ can be obtained as $J_K(k) = H_K(k)$ with $\phi' = 0$, which corresponds exactly to the form of the Kitaev lattice model. In the methods section, we give the detailed demonstration of such a correspondence. However, in the following analysis, we consider $J$ instead of $J_K(k)$. This is mainly due to two reasons. On the one hand, $J$ contains all the information of $J_K(k)$. The initial state and evolution equation of $J$ can be derived with the help of $J_K(k)$ (see methods); On the other hand, it is convenient to implement experimentally. In contrast to $J_K(k)$, which has complex parameters, $J$ only has real parameters. In this case, all coupling terms in $J$ can be controlled by only regulating the resistances.

Based on such correspondence, we can study Majorana-like edge states in the Kitaev model using circuits. Because the resistance $R_c$ in the circuits corresponds to the parameter $\mu$ in $H_K(k)$, we can generate a circuit segment with topological or trivial phase by only controlling $R_c$. If $4R_c < R_b$, the circuit segment has the trivial phase. If there is no intracell coupling, it has the topological phase. For the circuit network with topological and trivial phases, a topological edge state can emerge at the boundary between the trivial and the topological segment by feeding an external current to the network. By switching on the intracell coupling in the adjacent unit cell, a topological edge state is shifted to the adjacent unit cell. That is, we can move a topological edge state freely along the circuit network.

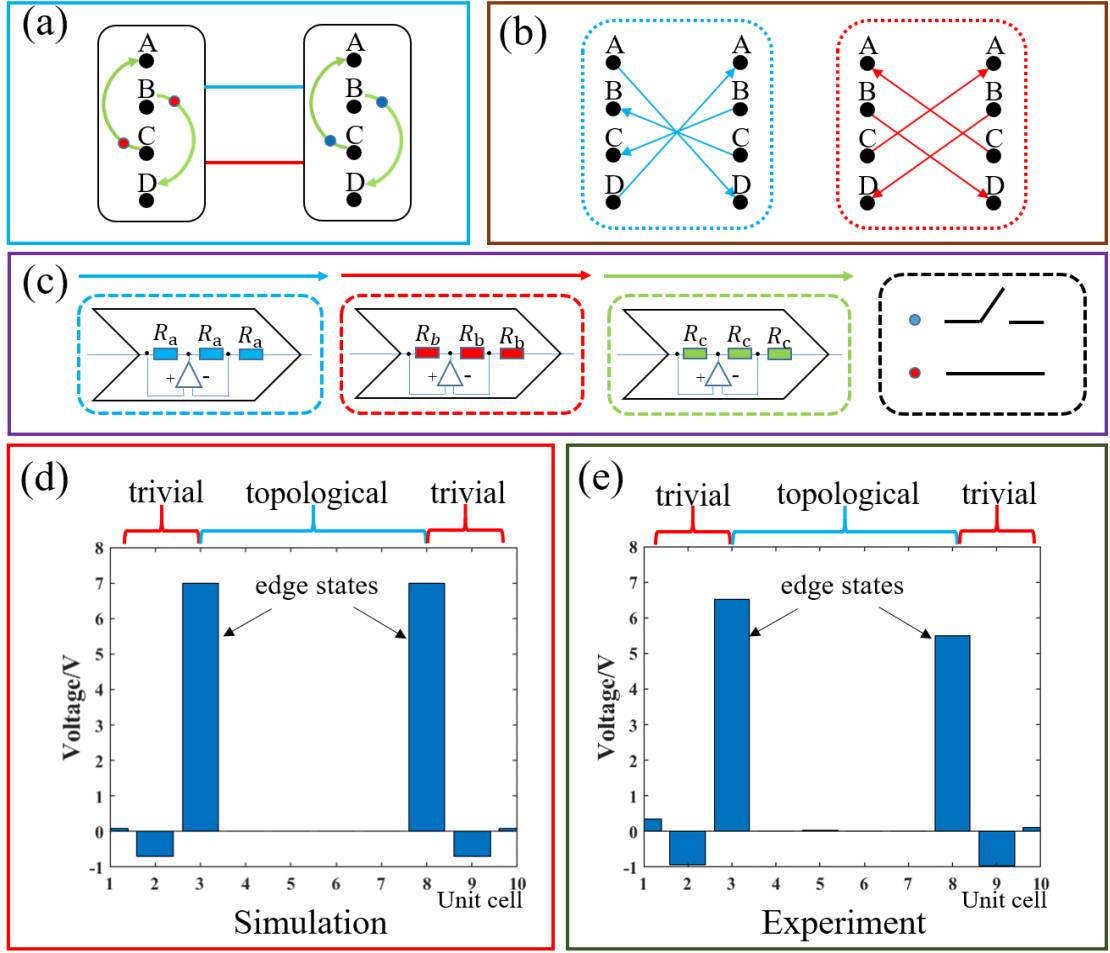

Figure 1 The circuit design and results for the Kitaev chain. (a) The designed electric circuit network for Kitaev chain. (b) The details of intercell couplings. (c) The schematic diagrams for the construction of the arrows. (d) The voltages of the whole chain with 10 units in simulation which are 0.01V, -0.7V, 7V, 0V, 0V, 0V, 0V, 7V, -0.7V and 0.01V from unit 1 to 10. A pair of topological edge states emerge at unit cells 3 and 8. Here $R_a = 1k\Omega$, $R_b = 1k\Omega$, $R_c = 100\Omega$. Operational ampliffifier is LT1013. (e) The experimental voltage distribution in the chain which are 0.36V, -0.94V, 6.52V, 0V, 0.04V, 0V, 0V, 5.5V, -0.95V and 0.12V from unit 1 to 10.

For example, we consider an electric circuit network with 10 units, which are labeled from 1 to 10. From 3 to 8, all units have topological phases, the units from 1 to 2 and 9 to 10 have trivial phases as shown in Fig. 1(d) (see top marks). To see the evolution of electrical signals in such a structure, we input positive voltage to node C in unit cell 3, and nodes A and B in unit cell 8. The negative voltage is input to node D in unit cell 3. The rest of the nodes are connected to the ground. The voltage excitations correspond to the initial state in the lattice model. Detailed discussion is given in the methods. After the voltages are excited, we immediately disconnect the input voltages simultaneously and measure the voltage distributions at various nodes. The simulation results are shown in Fig. 1(d). Here the evolution $6ms$ is taken before the signal decays. It is seen clearly that a pair of topological edge states emerge at the edge of the topological segment (units 3 and 8), which corresponds well to the distribution of edge states in the lattice model. Furthermore, we fabricate the corresponding circuit network, which its photograph is shown in Fig. S3 of Supplementary Note 3. Fig. 1(e) shows the experimental result of voltage distribution. For comparison, we find that the experiment and simulation results are basically identical. The differences between them come from the loss of the circuit and the error of components. Furthermore, to eliminate the influence of initial voltage on voltage distributions, we also input corresponding initial voltages to unit cells 2 and 9, the results of voltage distributions are exhibited in Supplementary Note 4. The edge states also emerge at the edge of the topological segment. This means that Majorana-like edge states are successfully observed in our designed circuit.

Based on the above edge states, we can design a braiding process by employing a T junction[17] as shown in the left panel of Fig. 2(a). The T junction provides the simplest wire network that enables the meaningful adiabatic exchange of Majorana-like zero modes. It has three legs (1-2, 2-3, and 2-4) made of Kitaev chains. The blue and brown colors correspond to topological and trivial segments, respectively. Such a T junction model can also be simulated by designing the circuit. The designed circuit network is shown in the red box of Fig. 2(a). The marks (1, 2, 3, and 4) in the circuit network correspond to those in the T junction model one by one. Legs 1-2 and 2-3 have the same structure as the circuit network in Fig. 1(a), in which the circuit phase is taken as $\phi=0$. The leg (2-4) has a variable phase $\phi$. In the leg (2-4), there are four units and three kinds of intercell couplings (purple, red and brown lines). The couplings marked by red are identical to those in Fig. 1(b). The detailed connections for the other couplings (purple and brown lines) are shown in the right panel of Fig. 2(a). Their corresponding structures of arrows are given in Fig. 2(b). Unlike the others, the connection represented by the purple arrow contains two INICs. This is because they correspond to variable resistances $R_a \cos^{-1}\phi$ with $\phi$ from 0 to $\pi$. When the $\phi$ changes from 0 to $\pi/2$, the upper INIC works. For the case from $\pi/2$ to $\pi$, the lower INIC works. In addition, proper grounding elements should be connected to each node. Details of the grounding parts are shown in Supplementary Note 5. For such a circuit network, we can also demonstrate that its Laplacian corresponds to the Hamiltonian of Kitaev chain with a variable phase $\phi$, which a detailed description is also given in the methods section. Furthermore, we also fabricate the corresponding circuit network, whose photograph is shown in Fig. S6 of Supplementary Note 6.

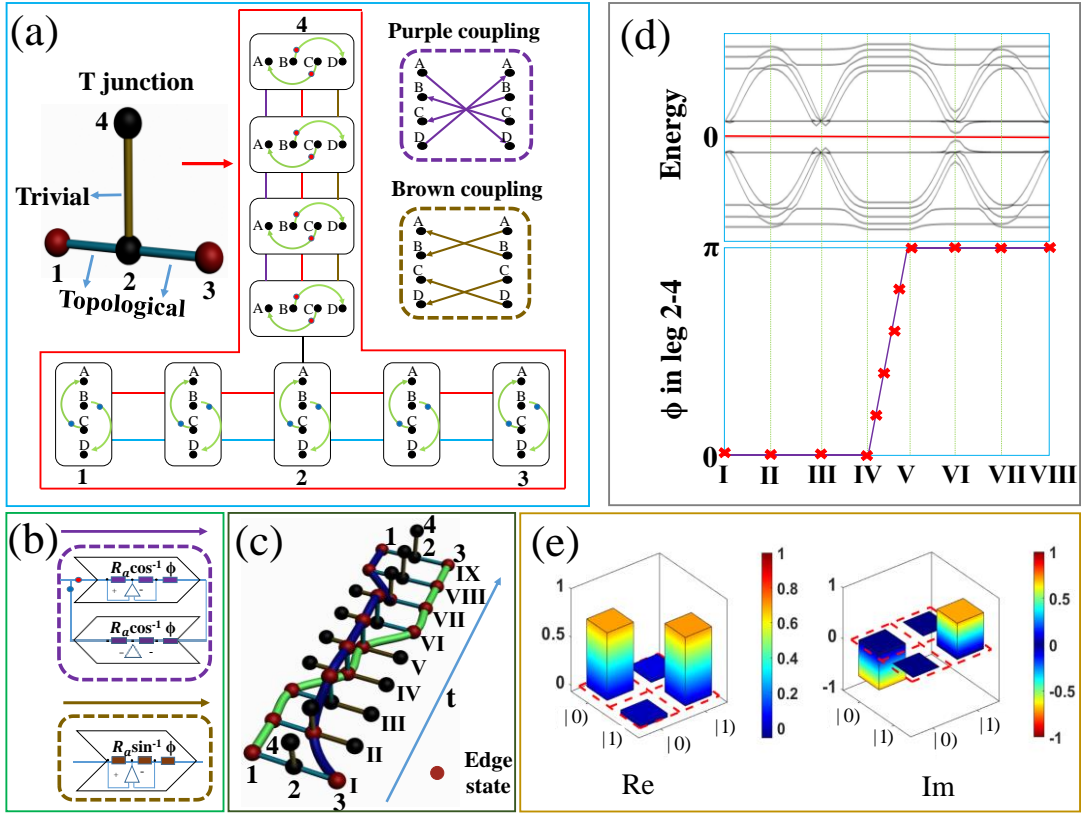

Figure 2 The circuit design and results for the braiding process. (a) Illustrations of an electric circuit for a T junction. The right panel shows the detailed connections of intercell couplings (purple and brown). The red and blue intercell couplings are identical to Fig. 1(b). (b) The schematic diagrams for the construction of the arrows. (c) The diagram of the braiding process with eight steps from I to IX. The red and black balls are the ends of the three legs in T-junction (1, 2, 3 and 4), where red balls represent the appearance of topological edge states at the ends of the legs, while black balls correspond to the case of topological trivial phase. By regulating the switches in resistances, the edge states evolve with time. (d) Evolution of the energy and the variation of phase in eight steps from I to IX. We choose $R_a = 1k\Omega$, $R_b = 1k\Omega$, $R_c = 100\Omega$. (e) The intensity of the braiding matrix.

From the circuit T junction, the designed braiding process is shown in Fig. 2(c). The braiding process which contains state transfer and phase rotation for two edges includes eight steps from I to IX. Before starting the braiding process, we set the initial state by regulating the switches in intracell couplings. The legs from 1 to 2 and 2 to 3 have topological phases. From 2 to 4, the leg has a trivial phase. The edge states appear at 1 and 3. In addition, we set $\phi = 0$ in all legs. Then we make the edge states evolve with time by regulating the switches. For example, in the step from I to II, the edge state moves from 3 to 2 by gradually closing the switches on the leg 2-3. In such a case, leg 1-2 (2-4) remains topological (trivial) segment, but leg 2-3 becomes a trivial segment. The edge states appear at 1 and 2. In the step from IV to V, the phase of leg 2-4 is from $\phi = 0$ to $\phi = \pi$. In the steps from I to IV and V to VIII, edge states transfer with time. From IV to V and VIII to IX, the phase rotation is performed. A detailed description of all steps and voltage evolution can be seen in the methods section. After eight steps, the braiding process is achieved with the result $V_1(0) \to V_3(0)$ and $V_3(0) \to -V_1(0)$. Here $V_i$ ($i = 1, 3$) represent voltages at the marked positions (1 and 3) in the T junction. The green and purple cylindrical lines in Fig. 2(c) display recorded positions of edge states at each step, where the transfer (braiding) process for two edge states is clearly shown as a function of time.

To quantitatively describe the above braiding process, we calculate the energy eigenvalue of the circuit T junction in eight evolution steps. The results are plotted in the upper panel of Fig. 2(d). It shows that the energy of the edge states remains zero during the process and they are well separated from all other states, which means the adiabatic evolution of the whole braiding process. The most important property of the topology is that it is robust against perturbation. To demonstrate this effect, we construct 10% coupling randomness. The corresponding energy eigenvalue is shown in Supplementary Note 7. The energy of the edge states is also well separated from the bulk spectrum, which indicates that the braiding process is robust against perturbation. In addition, in the down panel of Fig. 2(d) we plot $\phi$ as a function of evolution steps. The purple line and the red marks are theoretical and experimental results, respectively. The phase rotation from $\phi = 0$ to $\phi = \pi$ is observed in the steps from IV to V, which corresponds to the case of state transfer as shown in Fig. 2(c).

To better describe the braiding process, we measure the braiding matrix. The braiding matrix has the form $U_{\gamma_i \gamma_j} = \begin{pmatrix} e^{-i\pi/4} & 0 \\ 0 & e^{i\pi/4} \end{pmatrix}$ in the qubit representation, which indicates the braiding for zero modes $\gamma'_i$ and $\gamma'_j$. It exhibits a function of a $\pi/2$ phase gate. In Supplementary Note 8, we give a detailed description of the braiding matrix under the framework of quantum theory. In fact, the corresponding braiding matrix can also be obtained in the circuit. If we define $V_1 + iV_3 \rightarrow |0)$ and $V_1 - iV_3 \rightarrow |1)$ to represent the initial states, which correspond to $|0\rangle$ and $|1\rangle$ in the qubit representation, the derived braiding matrix in the circuit corresponds exactly to the form of the matrix under the framework of quantum theory. The detailed derivation is given in Supplementary Note 9. Here, the states in the paper are described by a slightly modified version of the familiar bra-ket notation of quantum mechanics.

The real and imaginary parts of output voltages $V_{|0)}$ and $V_{|1)}$ in the experimental braiding matrix are shown in Fig. 2(e). $V_{|i)}(i=0,1)$ indicates the output voltage after the braiding matrix acts on the initial state $|i)$. The parameters in the experiment are identical with above in Fig. 1(e). It is clearly seen that indeed the input states for $|0)$ and $|1)$ are successfully transformed into $1/\sqrt{2}(1-i)|0)$ and $1/\sqrt{2}(1+i)|1)$, respectively. It corresponds to the result of $\pi/2$ phase gate. To further quantify the experimental result, we calculate the fidelity of the braiding matrix, which is expressed as $F_H = |(\phi|M_{th}M_{exp}|\phi)|^2$ with $M_{th}$ and $M_{exp}$ being the theoretical and experimental results for matrixes and $|\phi)$ is defined as state $|0)$ or $|1)$. A high fidelity with $F=0.98\pm0.0069$ is obtained, which clearly confirms that our implementation of the braiding process has a good performance.

## Single-qubit unitary operations

Based on the T junction and braiding process described above, we can perform basic gate operations for topological quantum computation. It has been demonstrated that the combination of some single-qubit and two-qubit gates can generate any unitary transformation[48, 49]. In this section, we firstly discuss the circuit realization of the single-qubit unitary operations. Hadamard (H) gate, X gate, and Z gate are the basic single-qubit gates. According to the theory of topological quantum computing, two pairs (four) of Majorana-like zero modes are required to construct a qubit. For example, the left panel in Fig. 3(a) shows the theoretical scheme of the

H gate based on four zero modes, which are marked as $\gamma'_1$, $\gamma'_2$, $\gamma'_3$ and $\gamma'_4$, respectively. The scheme exhibits the braiding process as shown by the red lines, which includes three exchange operations among $\gamma'_1$, $\gamma'_2$ and $\gamma'_3$. Such a scheme for the H gate can also be simulated by designing the electric circuit. The designed circuit is shown in the right panel of Fig. 3(a), in which two pairs (four) of edge states corresponding to Majorana-like zero modes $\gamma_1$, $\gamma_2$, $\gamma_3$ and $\gamma_4$ are used. The braiding of these edge states also needs three exchange operations using T junctions, which correspond to the case in the left panel of Fig. 3(a) one by one. Detailed theoretical descriptions of the correspondence between the quantum scheme and circuit theory for the H gate are shown in Supplementary Note 10. Besides, we find that two kinds of braiding processes are needed to construct an H gate operation: braiding two edges across a topological segment (from $\gamma_1$ to $\gamma_2$ and $\gamma_3$ to $\gamma_4$) and a trivial segment (from $\gamma_2$ to $\gamma_3$). The braiding across a topological segment is identical with the above in Fig. 2(c). In the methods section, we describe all steps of the braiding across a trivial segment.

Furthermore, we fabricate the corresponding circuit network to measure matrix of output voltages for the H gate in the experiment. The measured results for the real and imaginary parts of output voltages are given in Fig. 3(b). The parameters are identical with the above in Fig. 2(e). It is clearly seen that the input states for $|0\rangle$ and $|1\rangle$ are successfully transformed into $|0\rangle+|1\rangle$ and $|0\rangle-|1\rangle$, respectively, which corresponds to the result of the H gate in the quantum scheme. The fidelity for the measured matrix can be obtained as $F = 0.9875 \pm 0.0083$. Such a high fidelity further indicates that the function of the H gate is well implemented. Detailed experimental dates for the H gate are also given in Supplementary Note 10.

We also exhibit the braiding for the Z gate in Fig. 3(c). The theoretical scheme in the left panel indicates that two exchange operations are needed to braid the Z gate. The right panel of Fig. 3(c) shows the designed circuit, in which the braiding of edge states corresponds to the theoretical scheme. In addition, the results of the measured matrix for the Z gate in the experiment are shown in Fig. 3(d). It can be seen that the input states for $|0\rangle$ and $|1\rangle$ are successfully transformed into $|0\rangle$ and $-|1\rangle$, respectively, which the fidelity is

$F = 0.9410 \pm 0.0102$. This means that the Z gate is also successfully achieved in our circuit experiment. The detailed theoretical descriptions and experimental results for the Z gate are given in Supplementary Note 11. Besides, we also design and measure the X gate in the circuit. The experimental results also correspond well with the theoretical scheme. Details can be seen in Supplementary Note 12.

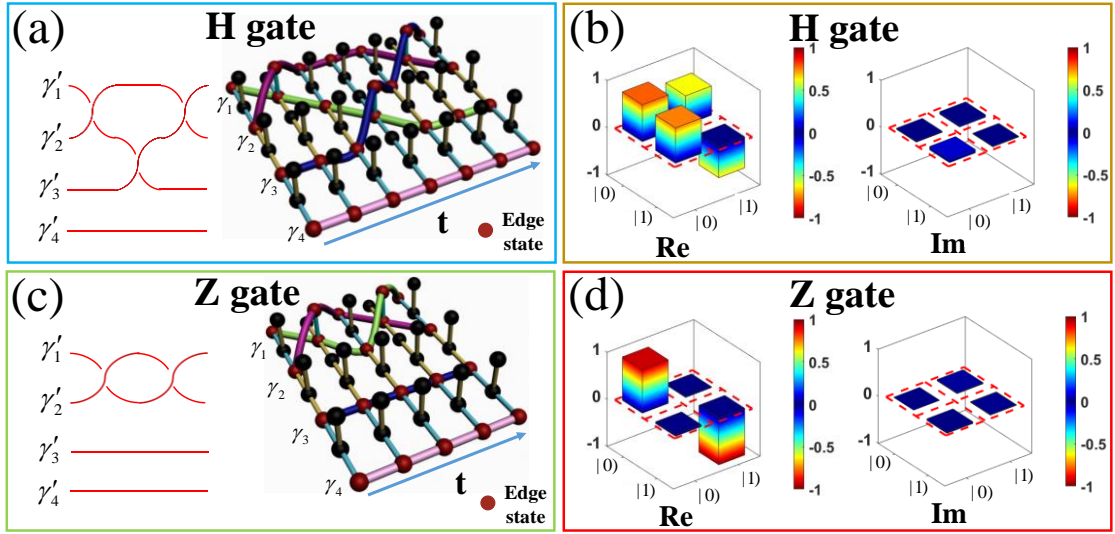

Figure 3 The circuit design and results for the single-qubit computations. The left panel in (a) and (c) illustrate the theoretical scheme for the H gate and the Z gate. The right panel in (a) and (c) show the circuit designs and transform of edge states. (b) and (d) are output voltages of measured matrices for the H gate and the Z gate, respectively.

**Two-qubit gates unitary operations**

We now provide the circuit design to construct two-qubit unitary operations. Two basic two-qubit gates are considered: the CNOT gate and the CZ gate. Three pairs (six) of Majorana-like zero modes are required to realize two-qubit operations. The left panel of Fig. 4(a) shows the theoretical scheme for the CNOT gate, in which the Majorana-like zero modes are marked as $\gamma'_1$, $\gamma'_2$, $\gamma'_3$, $\gamma'_4$, $\gamma'_5$ and $\gamma'_6$, respectively. The braiding process is shown by the red lines, which includes seven exchange operations. The designed circuit to simulate such a gate is shown in the right panel of Fig. 4(a), in which three pairs (six) of edge states corresponding to Majorana-like zero modes $\gamma_1$, $\gamma_2$, $\gamma_3$, $\gamma_4$, $\gamma_5$ and $\gamma_6$ are used. The braiding of these edge states corresponds to the theoretical scheme in the left panel in Fig. 4(a) one by one. Detailed theoretical descriptions of the correspondence between CNOT gate in the quantum scheme and circuit theory are shown in Supplementary Note 13.

The results of the measured output voltage matrix for the CNOT gate in the experiment are shown in Fig. 4(b). The parameters are identical with the above in Fig. 3(b). We use $V_{\gamma_1}+iV_{\gamma_2}\otimes V_{\gamma_5}+iV_{\gamma_6}\rightarrow|00)$, $V_{\gamma_1}+iV_{\gamma_2}\otimes V_{\gamma_5}-iV_{\gamma_6}\rightarrow|01)$, $V_{\gamma_1}-iV_{\gamma_2}\otimes V_{\gamma_5}+iV_{\gamma_6}\rightarrow|10)$ and $V_{\gamma_1}-iV_{\gamma_2}\otimes V_{\gamma_5}-iV_{\gamma_6}\rightarrow|11)$ to represent the initial states, which correspond to $|00\rangle$, $|01\rangle$, $|10\rangle$ and $|11\rangle$ in the quantum scheme. It is clearly seen that the input states for $|00)$, $|01)$, $|10)$ and $|11)$ are successfully transformed into $|00)$, $|01)$, $|11)$ and $|10)$ respectively. The fidelity of the CNOT gate can be obtained as $F=0.9089\pm0.0107$, indicating a nice performance of the CNOT function. Detailed experimental dates for the CNOT gate are also shown in Supplementary Note 13.

Fig. 4(c) displays the theoretical scheme and designed circuit for the CZ gate. The braiding of edge states in our designed circuit as shown in the right panel corresponds well to the theoretical scheme in the left panel, both of them include three exchange operations. From the experimental results of the measured output voltage matrix for the CZ gate, as shown in Fig. 4(d), it is seen clearly that the input states for $|00)$, $|01)$, $|10)$ and $|11)$ are successfully transformed into $|00)$, $|01)$, $|10)$ and $-|11)$, respectively. The fidelity for such an operation reaches $F=0.9409\pm0.0076$, indicating that the CZ gate can be successfully achieved in our circuit experiment. The detailed theoretical descriptions and experimental results for the CZ gate are given in Supplementary Note 14. After the single-qubit and two-qubit operations have been realized in the designed circuits, we next discuss how to perform quantum algorithm based on these gate operations.

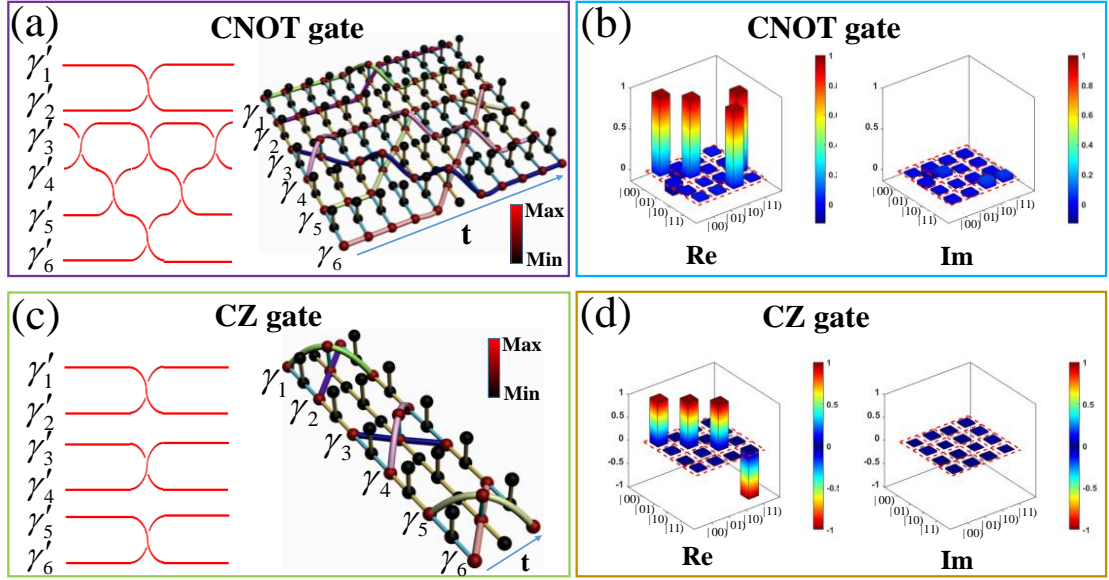

Figure 4 The circuit design and results for the two-qubit computations. The left panel in (a) and (c) illustrate the diagram of the CNOT gate and the CZ gate. The right panel in (a) and (c) show the braiding process. (b) and (d) are output voltage matrices of the CNOT gate and the CZ gate, respectively.

**Grover's search algorithm**

Grover's search algorithm is an important algorithm in quantum computation[26]. It is proved more efficient than the best classical algorithm and can solve difficult problems. The goal of Grover's search is to identify one out of $N$ elements of a database. The success probability to find the desired outcome is $1/N$ with one random guess in classical strategy. So, the average times of queries in finding the desired outcome is $N/2$. But, inputs of Grover's search can be processed simultaneously in superposition. It enhances the probability of finding the desired outcome in only $o(\sqrt{N})$, which shows quadratic speedup to the fastest classical algorithm in searching unsorted databases. Such an algorithm is extremely important, both from fundamental and practical standpoints. Grover's algorithm has been achieved in various systems[1-3]. Here, we demonstrate an implementation of Grover's fast quantum search using the braiding process in the circuit.

The universal quantum route diagram for the two-qubit Grover's search algorithm is shown in Fig. 5(a). It contains four parts: input, black box, inversion, and readout. The input qubits are prepared as $|0\rangle|0\rangle(|00\rangle)$. When they pass through two H gates, a superposition of four basis states $|00\rangle$, $|01\rangle$, $|10\rangle$ and $|11\rangle$ appears. One of which can be marked under the CZ and single-bit rotation gates ( $R_z(-a)(a=\alpha,\beta)$ ) in the black box, $\alpha$ and $\beta$ represent the rotation angles. If $\alpha\beta$ are taken as $00$, $0\pi$, $\pi 0$ and $\pi\pi$, the corresponding marks in the black box are $|00\rangle$, $|01\rangle$, $|10\rangle$ and $|11\rangle$, respectively. In principle, the marked state remains hidden. The goal of Grover's search is to identify the marked state by the inversion part. The inversion part inverts the amplitudes for each state about the mean value which can amplify the labelled amplitude and reduce the rest. After it, the marked state can be readout with probability 1 in theory. It indicates that Grover's algorithm needs only one calculation to identify the hidden marked state in the black box, whereas classically three evaluations are needed in the worst case, and 2.25 evaluations are needed on average.

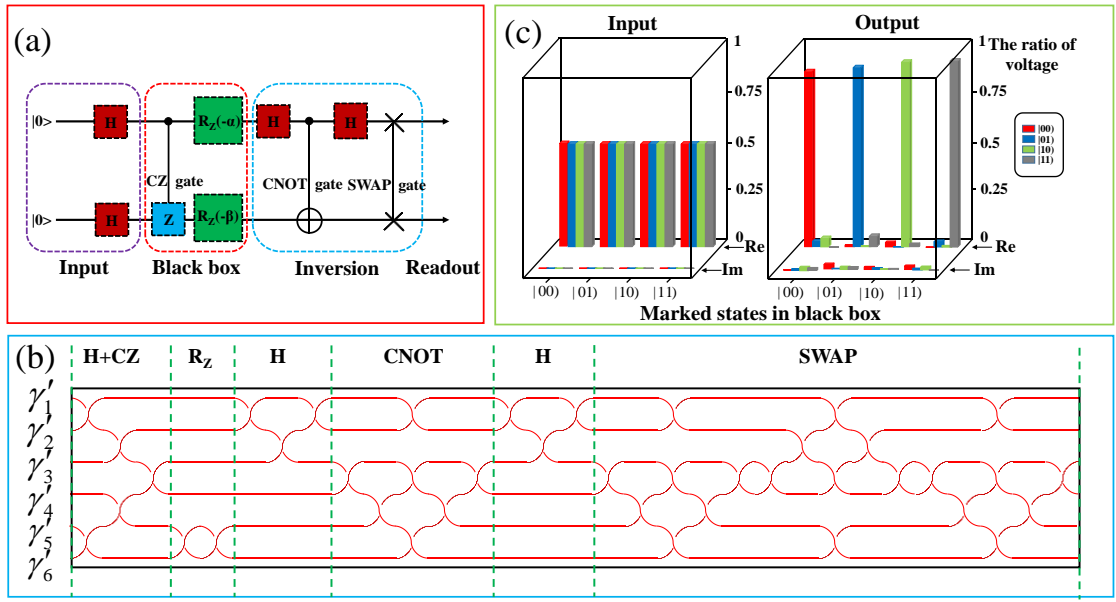

Figure 5 The circuit design and results for Grover's search algorithm. (a) The quantum diagram implementing Grover's search algorithm for two qubits. (b)The theoretical scheme of Grover's algorithm with $\alpha\beta=0\pi$. (c) The inputs and outputs of Grover's algorithm. The symbols "Re" and "Im" represent real and imaginary parts of $|00\rangle$, $|01\rangle$, $|10\rangle$ and $|11\rangle$ which are represented as the marked state in black box, respectively.

The corresponding braiding process we designed for the two-qubit Grover's search algorithm is shown in Fig. 5(b) with $\alpha\beta=0\pi$. Other cases ( $00$, $\pi 0$ and $\pi\pi$ ) are described in Supplementary Note 15. Such a scheme for the topological quantum computation corresponds to the universal quantum route diagram as shown in Fig. 5(a) one by one. The difference is that six Majorana-like zero modes ($\gamma_1'$, $\gamma_2'$, $\gamma_3'$, $\gamma_4'$, $\gamma_5'$ and $\gamma_6'$) and braiding process are used in the topological quantum computation scheme. The detailed theoretical descriptions for such a scheme are also given in Supplementary Note 15.

Furthermore, we design and fabricate circuit networks to implement such a scheme. In Supplementary Note 16, we provide the photograph of designed and fabricated circuits to simulate the scheme as shown in Fig. 5(b). Our fabricated circuit corresponds to the theoretical scheme. We use five T junctions to construct three pairs (six) of Majorana-like zero modes. By regulating the switches of resistances, we achieve the function of braiding process for edge states. After the whole braiding process, we measure the voltage of each node in T junctions. Using these voltages, we obtain the evolution in the circuit for Grover's search algorithm based on the braiding theory in Supplementary Note 9. When they act on the initial state after normalization, the output is in Fig. 5(c) for various marked states of Grover's search algorithm. Here, the parameters are taken identically with those in Fig. 4(b). The corresponding circuit theory and experimental dates are given in Supplementary Note 17. The bottom in Fig. 5(c) marks four states, $|00\rangle$, $|01\rangle$, $|10\rangle$ and $|11\rangle$, which correspond to the cases with $\alpha\beta = 00$, $0\pi$, $\pi 0$ and $\pi\pi$ in the black box, respectively. Different colors of square cylinders represent the ratio of the four states in the output voltage with only one evolution. For example, if we set $\alpha\beta = 00$, which indicates that the marked hidden state is $|00\rangle$, the four square cylinders in the input are equally high which indicate the superposition of four basis states $|00\rangle$, $|01\rangle$, $|10\rangle$ and $|11\rangle$. But the red square cylinder that represents the output probability of $|00\rangle$ is extremely high than the others. It clearly shows the amplitude amplification of the marked state. So, we need only one evolution to identify the hidden marked state in the black box, whereas classically three evaluations are needed in the worst case, and 2.25 evaluations are needed on average. In addition, the ratio of the correct outcome is above 90% in the present cases. These high-fidelity results constitute, to our knowledge, the first demonstration of a quantum search algorithm in a topological quantum computer.

**Discussion and conclusion**

We have provided a new circuit scheme to realize topological quantum computation.

Based on such a scheme, Majorana-like edge states have not only been observed experimentally, but also T junctions have been constructed for the braiding process. Furthermore, a set of one- and two-qubit unitary operations has been realized experimentally by the braiding operations. Our final realization of Grover's search algorithm strongly underlines the feasibility of such a scheme for topological quantum computation.

The reason why we can design the circuit to implement the scheme of the topological quantum computation is that there is a good correspondence between the circuit Laplacian and lattice Hamiltonian. Based on such correspondence, various topological physics and Schrödinger's equation have been simulated[36-47]. In fact, qubits, unitary transformation, superposition, and entanglement required by quantum computation can also be simulated using the circuit, as we have shown. Compared with other schemes, circuit networks possess remarkable advantages of being versatile and reconfigurable, and classical circuit technology is relatively mature. If topological quantum computation can be realized using electric circuits, it is expected to avoid some problems faced by the present quantum computation schemes, such as decoherence and scalability. In addition, the energy loss caused by resistances in the circuit network can also be supplemented by connecting to the power. That is, the constructed system can work stably. Thus, our work paves the way for the construction of a practical and robust fast information processing system to serve society.

**Methods**

**The correspondence between the Kitaev lattice and our designed circuit**

In this section, we give a detailed correspondence of the eigen-equation between the Kitaev lattice model and our designed circuit. Because the circuit structure in Fig. 1(a) is a special case of Fig. 2(a) with $\phi=0$, in the following we discuss the correspondence based on the circuit structure in Fig. 2(a). According to Kirchhoff's law, the relation between the node current and voltage should satisfy the following equation:

$$I_m = -C_g \frac{dV_m}{dt} - \frac{V_m}{R_g} + \sum_{\langle n \rangle} \frac{V_n - V_m}{R_{\mathrm{inter}}}, \qquad (2)$$

where $I_m$ and $V_m$ are the net current and voltage of node $m$. $V_n$ is the voltage of node $n$ which is connected to the node $m$. $R_g$ and $C_g$ are the resistance and capacitance between the node *m* and ground. The $R_{inter}$ is the resistance between the node *m* and *n*. <n> indicates that the summation is limited within connected nodes. According to the circuit in Fig. 2(a), we can write the currents that flow into each node as

$$I_A^n = -C\frac{dV_A^n}{dt} + \frac{V_C^{n-1} - V_A^n}{R_b} + \frac{V_D^{n-1} - V_A^n}{R_a \cos^{-1}\phi} + \frac{-V_A^n}{-R_c} + \frac{V_B^{n-1} - V_A^n}{R_a \sin^{-1}\phi}$$
$$+ \frac{V_D^{n+1} - V_A^n}{-R_a \cos^{-1}\phi} + \frac{V_C^{n+1} - V_A^n}{R_b} + \frac{V_C^n - V_A^n}{R_c} + \frac{V_B^{n+1} - V_A^n}{-R_a \sin^{-1}\phi} + \frac{-2V_A^n}{-R_b},$$

$$I_B^n = -C\frac{dV_B^n}{dt} + \frac{V_D^{n-1} - V_B^n}{-R_b} + \frac{V_C^{n-1} - V_B^n}{-R_a \cos^{-1}\phi} + \frac{-V_B^n}{R_c} + \frac{V_A^{n-1} - V_B^n}{R_a \sin^{-1}\phi}$$
$$+ \frac{V_C^{n+1} - V_B^n}{R_a \cos^{-1}\phi} + \frac{V_D^{n+1} - V_B^n}{-R_b} + \frac{V_D^n - V_B^n}{-R_c} + \frac{V_A^{n+1} - V_B^n}{-R_a \sin^{-1}\phi} + \frac{-2V_B^n}{R_b},$$

$$I_C^n = -C\frac{dV_C^n}{dt} + \frac{V_A^{n-1} - V_C^n}{-R_b} + \frac{V_B^{n-1} - V_C^n}{-R_a \cos^{-1}\phi} + \frac{-V_C^n}{R_c} + \frac{V_D^{n-1} - V_C^n}{R_a \sin^{-1}\phi}$$
$$+ \frac{V_B^{n+1} - V_C^n}{R_a \cos^{-1}\phi} + \frac{V_A^{n+1} - V_C^n}{-R_b} + \frac{V_A^{n+1} - V_C^n}{-R_c} + \frac{V_D^{n+1} - V_C^n}{-R_a \sin^{-1}\phi} + \frac{-2V_C^n}{R_b},$$

$$I_D^n = -C\frac{dV_D^n}{dt} + \frac{V_B^{n-1} - V_D^n}{R_b} + \frac{V_A^{n-1} - V_D^n}{R_a \cos^{-1}\phi} + \frac{-V_D^n}{-R_c} + \frac{V_C^{n-1} - V_D^n}{R_a \sin^{-1}\phi}$$
$$+ \frac{V_A^{n+1} - V_D^n}{-R_a \cos^{-1}\phi} + \frac{V_B^{n+1} - V_D^n}{R_b} + \frac{V_B^n - V_D^n}{R_c} + \frac{V_C^{n+1} - V_D^n}{-R_a \sin^{-1}\phi} + \frac{-2V_D^n}{-R_b}, \quad (3)$$

where $I_i^n$ and $V_i^n$ are the current and voltage of the node *i* in the unit *n*. We assume that there is no external source, so that the currents flow into each node are zero. In this case, Eq. (3) becomes

$$\frac{dV_A^n}{dt} = -\frac{1}{C}\left(\frac{V_D^{n+1} - V_D^{n-1}}{R_a \cos^{-1}\phi} - \frac{V_C^{n+1} + V_C^{n-1}}{R_b} - \frac{V_C^n}{R_c} + \frac{V_B^{n+1} - V_B^{n-1}}{R_a \sin^{-1}\phi}\right),$$

$$\frac{dV_B^n}{dt} = -\frac{1}{C}\left(-\frac{V_C^{n+1} - V_C^{n-1}}{R_a \cos^{-1}\phi} + \frac{V_D^{n+1} + V_D^{n-1}}{R_b} + \frac{V_D^n}{R_c} + \frac{V_A^{n+1} - V_A^{n-1}}{R_a \sin^{-1}\phi}\right),$$

$$\frac{dV_C^n}{dt} = -\frac{1}{C}\left(-\frac{V_B^{n+1} - V_B^{n-1}}{R_a \cos^{-1}\phi} + \frac{V_A^{n+1} + V_A^{n-1}}{R_b} + \frac{V_A^n}{R_c} + \frac{V_D^{n+1} - V_D^{n-1}}{R_a \sin^{-1}\phi}\right),$$

$$\frac{dV_D^n}{dt} = -\frac{1}{C}\left(\frac{V_A^{n+1} - V_A^{n-1}}{R_a \cos^{-1}\phi} - \frac{V_B^{n+1} + V_B^{n-1}}{R_b} - \frac{V_B^n}{R_c} + \frac{V_C^{n+1} - V_C^{n-1}}{R_a \sin^{-1}\phi}\right). \quad (4)$$

We can write Eq. (4) into the matrix form

$$\partial_t \begin{pmatrix} V_A^n \\ V_B^n \\ V_C^n \\ V_D^n \end{pmatrix} = J \begin{pmatrix} V_A^n \\ V_B^n \\ V_C^n \\ V_D^n \end{pmatrix}, \tag{5}$$

where

$$J = \frac{-1}{C} \begin{pmatrix} 0 & R_a^{-1}\sin\phi e^{ik} - R_a^{-1}\sin\phi e^{-ik} & -R_c^{-1} - R_b^{-1}e^{ik} - R_b^{-1}e^{-ik} & R_a^{-1}\cos\phi e^{ik} - R_a^{-1}\cos\phi e^{-ik} \\ R_a^{-1}\sin\phi e^{ik} - R_a^{-1}\sin\phi e^{-ik} & 0 & -R_a^{-1}\cos\phi e^{ik} + R_a^{-1}\cos\phi e^{-ik} & R_c^{-1} + R_b^{-1}e^{ik} + R_b^{-1}e^{-ik} \\ R_c^{-1} + R_b^{-1}e^{ik} + R_b^{-1}e^{-ik} & -R_a^{-1}\cos\phi e^{ik} + R_a^{-1}\cos\phi e^{-ik} & 0 & R_a^{-1}\sin\phi e^{ik} - R_a^{-1}\sin\phi e^{-ik} \\ R_a^{-1}\cos\phi e^{ik} - R_a^{-1}\cos\phi e^{-ik} & -R_c^{-1} - R_b^{-1}e^{ik} - R_b^{-1}e^{-ik} & R_a^{-1}\sin\phi e^{ik} - R_a^{-1}\sin\phi e^{-ik} & 0 \end{pmatrix}. \tag{6}$$

Eq. (6) can further be written as Eq. (1). If we choose the parameters to be $(R_aC)^{-1} = \frac{\Delta}{4}$, $(R_bC)^{-1} = \frac{t}{4}$ and $(R_cC)^{-1} = \frac{\mu}{2}$. Then, $J_K(k)$ can have the form as $H_K(k)$, which corresponds exactly to the form of the Kitaev lattice model. In the circuit of Fig. 1(a), red and blue couplings only exist. So, the circuit Laplacian can be written as

$$J = \frac{-1}{C} \begin{pmatrix} 0 & 0 & -R_c^{-1} - R_b^{-1}e^{ik} - R_b^{-1}e^{-ik} & R_a^{-1}e^{ik} - R_a^{-1}e^{-ik} \\ 0 & 0 & -R_a^{-1}e^{ik} + R_a^{-1}e^{-ik} & R_c^{-1} + R_b^{-1}e^{ik} + R_b^{-1}e^{-ik} \\ R_c^{-1} + R_b^{-1}e^{ik} + R_b^{-1}e^{-ik} & -R_a^{-1}e^{ik} + R_a^{-1}e^{-ik} & 0 & 0 \\ R_a^{-1}e^{ik} - R_a^{-1}e^{-ik} & -R_c^{-1} - R_b^{-1}e^{ik} - R_b^{-1}e^{-ik} & 0 & 0 \end{pmatrix}, \tag{7}$$

which is Eq. (1) with $\phi = 0$.

### The evolution and initial voltage excitation of the circuit

We first discuss the evolving relationship between the circuit and lattice model. The evolution of the circuit's initial state $V_0(0)$ with time satisfies the equation:

$$V_0(T) = e^{Jt}V_0(0), \tag{8}$$

where $J$ is identical with Eq. (1). We make a similar transformation on the matrix

$$J_1 = P^{-1}JP = \begin{bmatrix} -iJ_K(k) & 0 \\ 0 & iJ_K^*(k) \end{bmatrix}, \tag{9}$$

where $P = \frac{1}{\sqrt{2}}\begin{bmatrix} I & iI \\ iI & I \end{bmatrix}$ and $P^{-1} = \frac{1}{\sqrt{2}}\begin{bmatrix} I & -iI \\ -iI & I \end{bmatrix}$. So that the evolution of $J_1$ satisfies

$$V_1(T) = e^{-J_1 t}V_1(0) = e^{-P^{-1}JPt}V_1(0) = P^{-1}e^{-Jt}PV_1(0) \Rightarrow PV_1(T) = e^{-Jt}PV_1(0). \tag{10}$$

The relationships of the initial state and the finial state between $J$ and $J_1$ can be written as

$$PV_1(0) = V_0(0),$$

$$PV_1(T) = V_0(T). \tag{11}$$

In order to correspond with Hamiltonian in the lattice model, we set the initial state of $J_1$ to be

$$V_1(0) = \begin{pmatrix} V_2(0) \\ -iV_2^*(0) \end{pmatrix}, \tag{12}$$

where $V_2(0)$ is the initial state of $J_K(k)$. We make $V_2(T)$ be the projection of $V_1(T)$ on $(1 \ 0) \otimes I_N$, and get

$$\begin{aligned} V_2(T) &= (1 \ 0) \otimes I_N V_1(T) \\ &= (1 \ 0) \otimes I_N \cdot e^{J_1 t} \cdot \begin{pmatrix} V_2(0) \\ -iV_2^*(0) \end{pmatrix}, \\ &= e^{-iJ_K(k)t} V_2(0), \end{aligned} \tag{13}$$

which corresponds to the evolution of the wave function in the lattice model.

We next discuss the initial state of the design circuit. It can be seen in Eq. (1) that $J$ contains real and imaginary parts of $J_K(k)$. After combining Eq. (11) and (12), we get $V_0(0) = \frac{1}{\sqrt{2}} \begin{pmatrix} V_2(0) + V_2^*(0) \\ iV_2(0) - iV_2^*(0) \end{pmatrix}$. So, voltages in nodes A and B (C and D) in the designed circuit are the real (imaginary) parts of voltages in $H_K(k)$. We define the initial voltage $V_0(0)$ be

$$V_0^A(0) = \begin{pmatrix} a_1 \\ b_1 \end{pmatrix},$$

$$V_0^B(0) = \begin{pmatrix} a_2 \\ b_2 \end{pmatrix}, \tag{14}$$

where a (b) is the real (imaginary) part of the initial voltage. According to Eq. (11) and (12), we have

$$V_1^A(0) = \begin{pmatrix} \dfrac{a_1 - ib_1}{2} \\ \dfrac{-ia_1 + b_1}{2} \end{pmatrix} = \begin{pmatrix} V_2^A(0) \\ -iV_2^{A*}(0) \end{pmatrix},$$

$$V_1^B(0) = \begin{pmatrix} \dfrac{a_2 - ib_2}{2} \\ \dfrac{-ia_2 + b_2}{2} \end{pmatrix} = \begin{pmatrix} V_2^B(0) \\ -iV_2^{B*}(0) \end{pmatrix}. \tag{15}$$

The initial state of $J_K(k)$ is identical with Hamiltonian $H_K(k)$:

$$V_2^A(0) = (-i, i, 0, \cdots, 0)^T,$$

$$V_2^B(0) = (0, \cdots, 0, 1, 1)^T. \tag{16}$$

So, we get the initial state of the design circuit

$$V_0^A(0) = \frac{1}{2}\begin{pmatrix} \Theta \\ (1, -1, 0, \cdots, 0)^T \end{pmatrix},$$

$$V_0^B(0) = \frac{1}{2}\begin{pmatrix} (0, \cdots, 0, 1, 1)^T \\ \Theta \end{pmatrix}, \tag{17}$$

where $\Theta = (0, \cdots, 0)^T$. In our designed circuit, nodes A and B (C and D) are the real (imaginary) parts of the signal. So, for the circuit with 10 units. We input positive voltage to node C in unit cell 3 and nodes A and B in unit cell 8. We input negative voltage to node D in unit cell 3.

**Detailed description on the braiding process**

The whole braiding process can be achieved by eight steps from I to IX as shown in Fig. 2(c). We now explain the eight steps from I to IX in detail:

[(1): I→II] We move the edge on 3 toward 2 by gradually closing the switches on leg 2-3.

[(2): II→III] When the edge reaches 2, we turn the trivial segment on leg 2-4 topological gradually by opening the switches so that the edge moves upward.

[(3): III→IV] When leg 2-4 becomes topological entirely, we move the edge on 1 toward 2.

[(4): IV→V] When the edge reaches 2, we rotate the phase of leg 2-4 from $\phi = 0$ to $\phi = \pi$.

[(5): V→VI] When the phase of leg 2-4 becomes $\phi = \pi$, we move the edge on 2 toward 3.

[(6): VI→VII] When leg 2-3 becomes topological entirely, we move the edge on 4 toward 2.

[(7): VII→VIII] When the edge reaches 2, we move it to 1.

[(8): VIII→IX] When leg 1-2 becomes topological entirely, we rotate the phase of leg 2-4 from $\phi = \pi$ to $\phi = 0$.

A Detailed description of phase rotation in Steps IV to V and VIII to IX can be seen in

Supplementary Note 18. When it comes to gate operations, the braiding across a trivial segment is needed. The designed braiding process is shown in Supplementary Note 19. We now explain the eight steps of it in detail:

[(1): I→II] We move the edge on 7 toward 6 by gradually opening the switches on leg 6-7.

[(2): II→III] When the edge reaches 6, we turn the trivial segment on leg 6-8 topological gradually by opening the switches so that the edge moves upward.

[(3): III→IV] When leg 6-8 becomes topological entirely, we move the edge on 5 toward 6.

[(4): IV→V] When the edge reaches 6, we rotate the phase of leg 6-8 from $\phi=0$ to $\phi=\pi$.

[(5): V→VI] When the phase of leg 6-8 becomes $\phi=\pi$, we move the edge on 6 toward 7.

[(6): VI→VII] When leg 6-7 becomes trivial entirely, we move the edge on 8 toward 6.

[(7): VII→VIII] When the edge reaches 6, we move it to 5.

[(8): VIII→IX] When leg 1-2 becomes trivial entirely, we rotate the phase of leg 6-8 from $\phi=\pi$ to $\phi=0$.

We next numerically analyze the voltage evolution of the braiding process. For a circuit network with N unit cells in real space, we choose two eigenstates as the initial voltages of the system

$$V_1(0) = \begin{pmatrix} (1,-1,0,\cdots,0)^T \\ \Theta \end{pmatrix},$$

$$V_3(0) = \begin{pmatrix} \Theta \\ (0,\cdots,0,1,1)^T \end{pmatrix}, \qquad (18)$$

where $\Theta=(0,\cdots,0)^T$. After the braiding process I to IX, we find that these two eigenstates satisfied $V_1(0) \to V_3(0)$ and $V_3(0) \to -V_1(0)$. It has the same form as the non-Abelian statistics which is realized by the braiding process of Majorana-like zero modes. In fact, we can calculate the evolution of two edge states in the braiding process. For a time-dependent Schrödinger equation in the circuit

$$J(t)V_\alpha(t) = \partial_t V_\alpha(t). \qquad (19)$$

If the evolution is slow enough, the adiabatic approximation can guarantee

$$V_\alpha(t) = \sum_\beta U_{\beta\alpha} V_\beta(t). \qquad (20)$$

$V_\beta(t)$ satisfies the time-independent Schrödinger equation

$$J(t)V_\beta(t) = \varepsilon(t)V_\beta(t), \tag{21}$$

where $\varepsilon$ is the eigenvalue of the system. $U_{\beta\alpha}$ is the element of multi-level Berry phase matrix

$$U = P\exp\left(i\int_\Gamma A d\lambda\right), \tag{22}$$

where A is Berry connection matrix,

$$A_{\beta\alpha} = iV_\beta^\dagger(t)\partial\lambda V_\alpha(t). \tag{23}$$

Berry phase matrix $U$ is only related to $\Gamma(\lambda)$. For our braiding process I to IX, we can divide the path into k parts. The phase matrix can be expressed as

$$U = W_k W_{k-1} \cdots W_2 W_1, \tag{24}$$

where $(W_j)_{\beta\alpha} = V_\beta(\lambda_{j+1})V_\alpha(\lambda_j)$. Since our braiding process is a closed path, the corresponding U is gauge-invariant. For our braiding process, we can get

$$U = \begin{pmatrix} 0 & -1 \\ 1 & 0 \end{pmatrix}. \tag{25}$$

Combined Eq. (20), we get the evolution results $V_1(0) \to V_3(0)$ and $V_3(0) \to -V_1(0)$.

**Data availability.** The data that support the plots within this paper are available from the corresponding author upon reasonable request.

**Code availability.** The code that supports the plots within this paper is available from the corresponding author upon reasonable request.

48. Deutsch, D., Barenco, A. & Ekert, A. Universality in Quantum Computation. *Proc. R. Soc. Lond. A* **449,** 669–677 (1995).

49. Barenco, A. et al. Elementary gates for quantum computation. *Phys. Rev. A* **52,** 3457–3467 (1995).

**Acknowledgements**
This work was supported by the National Key R & D Program of China under Grant No. 2022YFA1404904 and the National Natural Science Foundation of China (No. 12234004).

**Author contributions.** D. Y. Zou, N. Q. Pan and T. Chen finished the theoretical scheme and designed the circuit simulator. D. Y. Zou finished the experiments with the help of H. J. Sun. D. Y. Zou and X. D. Zhang wrote the manuscript with the help of N. Q. Pan and T. Chen. X. D. Zhang initiated and designed this research project.

**Competing interests**
The authors declare no competing interests.

# Supplementary Information for

# Experimental topological quantum computing with electric circuits

Deyuan Zou [1,*], Naiqiao Pan[*], Tian Chen[1,*], Houjun Sun[2], and Xiangdong Zhang[1,+]

[1] Key Laboratory of advanced optoelectronic quantum architecture and measurements of Ministry of Education, Beijing Key Laboratory of Nanophotonics & Ultrafine Optoelectronic Systems, School of Physics, Beijing Institute of Technology, 100081, Beijing, China

[2] Beijing Key Laboratory of Millimeter wave and Terahertz Techniques, School of Information and Electronics, Beijing Institute of Technology, Beijing 100081, China

*These authors contributed equally to this work. [+$]Author to whom any correspondence should be addressed. E-mail: zhangxd@bit.edu.cn, chentian@bit.edu.cn

Supplementary Note 1. Correspondence between topological insulator model and p-wave superconducting chain model.
Supplementary Note 2. Groundings of nodes in the circuit network for Majorana-like edge states.
Supplementary Note 3. The photograph of the fabricated Kitaev chain in the electric circuit.
Supplementary Note 4. The results of voltage distributions when we input voltages to unit cells 2 and 9.
Supplementary Note 5. Groundings of the T junction.
Supplementary Note 6. The photograph of the fabricated T junction in the electric circuit.
Supplementary Note 7. The energy eigenvalue with perturbation during the braiding process.
Supplementary Note 8. Braiding matrix under the framework of quantum theory.
Supplementary Note 9. The braiding matrix in the circuit.
Supplementary Note 10. The theoretical description and experimental data for H gate.
Supplementary Note 11. The theoretical description and experimental data for Z gate.
Supplementary Note 12. The theoretical description and experimental data for X gate.
Supplementary Note 13. The theoretical description and experimental data for CNOT gate.
Supplementary Note 14. The theoretical description and experimental data for CZ gate.
Supplementary Note 15. Theoretical scheme for Grover's search.
Supplementary Note 16. The photograph of the designed and fabricated circuit to achieve Grover's search.
Supplementary Note 17. Circuit theory and experimental dates for Grover's search.
Supplementary Note 18. A detailed description of phase rotation.
Supplementary Note 19. The braiding process of the two edges across a trivial segment.

## Supplementary Note 1. Correspondence between topological insulator model and p-wave superconducting chain model.

The simplest system capable of carrying MZM is the one-dimensional spin-free p-wave superconducting chain model. The Hamiltonian can be written as

$$H = -\mu \sum_{x=1}^{N} c_x^\dagger c_x - \frac{1}{2} \sum_{x=1}^{N-1} (t c_x^\dagger c_{x+1} + \Delta e^{i\phi} c_x c_{x+1} + \text{H.c.}) \ . \tag{S1}$$

It can define the Majorana operators

$$\gamma_{x,A} = i\left(c_x^\dagger e^{-i\phi/2} - c_x e^{i\phi/2}\right),$$

$$\gamma_{x,B} = c_x^\dagger e^{-i\phi/2} + c_x e^{i\phi/2} \ . \tag{S2}$$

If we set $\mu = 0, \Delta = t \neq 0$, the two Majorana at different positions are coupled with each other. So, there are two isolated Majorana modes at both ends of the fermion chain.

The momentum space Hamiltonian can be obtained by taking periodic boundary conditions for the above Hamiltonian and making Fourier transform $H(k) = \frac{1}{2} \sum C_k^\dagger \tilde{H}(k) C_k$, where

$$\tilde{H}(k) = \frac{1}{2} \begin{pmatrix} \varepsilon_k & i\Delta e^{-i\phi} \sin k \\ -i\Delta e^{i\phi} \sin k & -\varepsilon_k \end{pmatrix}, \tag{S3}$$

$C_k^\dagger = (c_k^\dagger, c_{-k})$ is Nambu operator.

In order to convert the above topological superconducting model into a topological insulator model, we make a replacement $c_k^\dagger \to A_k^\dagger$ and $c_{-k} \to B_k^\dagger$. **It means that each unit cell contains two sub-lattices $A$ and $B$. The real space Hamiltonian** of the topological insulator model **can be expressed as**

$$H = -\mu \sum_{x=1}^{N} A_x^\dagger A_x + \mu \sum_{x=1}^{N} B_x^\dagger B_x - \frac{1}{2} \sum_x (t A_x^\dagger A_{x+1} - t B_x^\dagger B_{x+1} + \Delta e^{-i\phi} A_x^\dagger B_{x+1} - \Delta e^{-i\phi} A_{x+1}^\dagger B_x + \text{H.c.}).$$

$$\tag{S4}$$

It is found that the **Hamiltonian** of the topological insulator model corresponds exactly to the **Hamiltonian** of the topological superconducting model in Eq. (S1). Therefore, the topological insulator model also has particle-hole symmetry. When $\mu = 0, \Delta = t \neq 0$, **it has zero-energy eigenvalues. These two degenerate zero-energy eigenstates can also correspond to two Majorana.**

**Supplementary Note 2. Groundings of nodes in the circuit network for Majorana-like edge states**

In order to obtain the circuit Laplacian $J$, proper grounding elements should be connected on each node. In this part, we show the groundings of each node in the circuit network. Fig.S1(a) shows the groundings in Fig. 1(a) in the main text. A-D are four nodes in a unit cell. Two columns of groundings correspond to trivial and topological cells, respectively. Different colors of graphs correspond to different INICs in Fig. 1(c) in the main text. Detailed structures are shown in Fig. S1(b). With the help of Fig. S1, we can obtain the grounding of a circuit network with any number of unit cells.

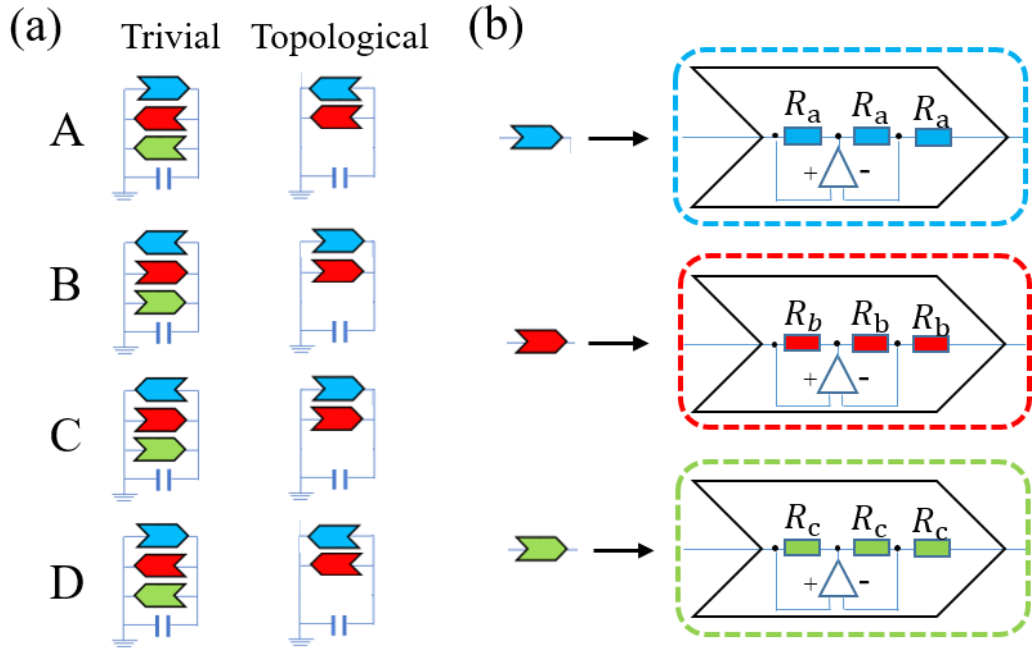

Figure S1 (a) The groundings of Kitaev model in Fig. 1(a). A-D are four nodes in a unit cell. (b) The detailed structures of different graphs.

In experiments, we consider an electric circuit network with 10 unit cells. Fig. S2 shows the groundings of Kitaev chain with 10 unit cells. ① and ⑩ are two edges of the chain, which have more grounding elements. ①-② and ⑨-⑩ are trivial segments, so they have green groundings. ③-⑧ are topological segments. They have the same groundings.

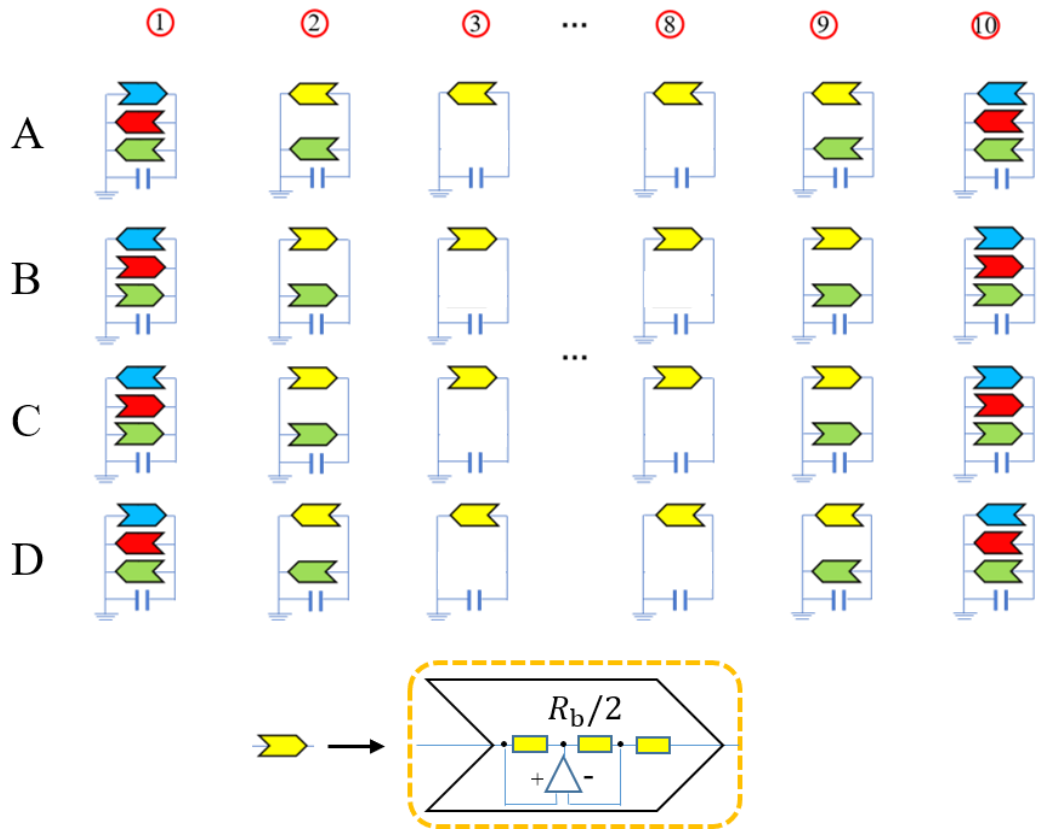

Figure S2 The groundings of Kitaev chain with 10 units. ①-⑩ represent 10 unit cells. The down panel is the structure of yellow INIC, and the values of the three resistors are $R_b/2$ .

**Supplementary Note 3. The photograph of the fabricated Kitaev chain in the electric circuit.**

The photograph of the fabricated Kitaev chain with 10 units is exhibited in Fig. S3. INIC is constructed with the help of opamp, for which the model is lt1013. A relay is used to disconnect the input voltages simultaneously. Besides, we use a Pin header to measure the voltage at various nodes.

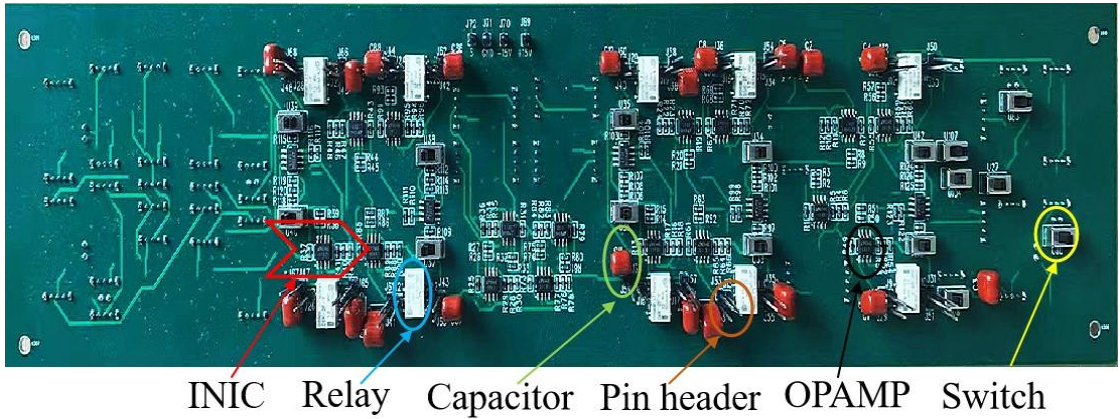

Figure S3 The photograph of Kitaev chain in the experiment.

## Supplementary Note 4. The results of voltage distributions when we input voltages to unit cells 2 and 9

Fig. S4 shows the results of voltage distributions when we input corresponding initial voltages to unit cells 2 and 9. It is clearly seen that the edge states also emerge at the edge of the topological segment (unit cells 3 and 8).

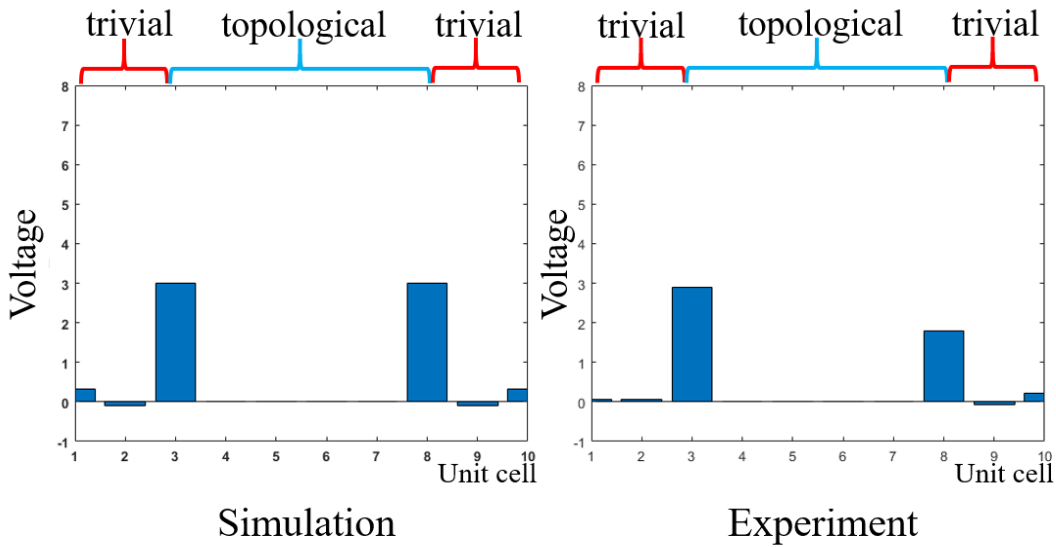

Figure S4 The results of voltage distributions when we input voltages to unit cells 2 and 9.

## Supplementary Note 5. Groundings of T junction

In order to obtain the circuit Laplacian, proper grounding elements should be connected on each node. Fig. S5(a) shows the groundings of the four unit cells from 2 to 4 in the T junction. For the case from 1 to 3, the groundings are identical with the circuit network in Fig. 1(a).

Different colors of graphs correspond to different INICs in Fig. 2(b). Detailed structures are shown in Fig. S5(b).

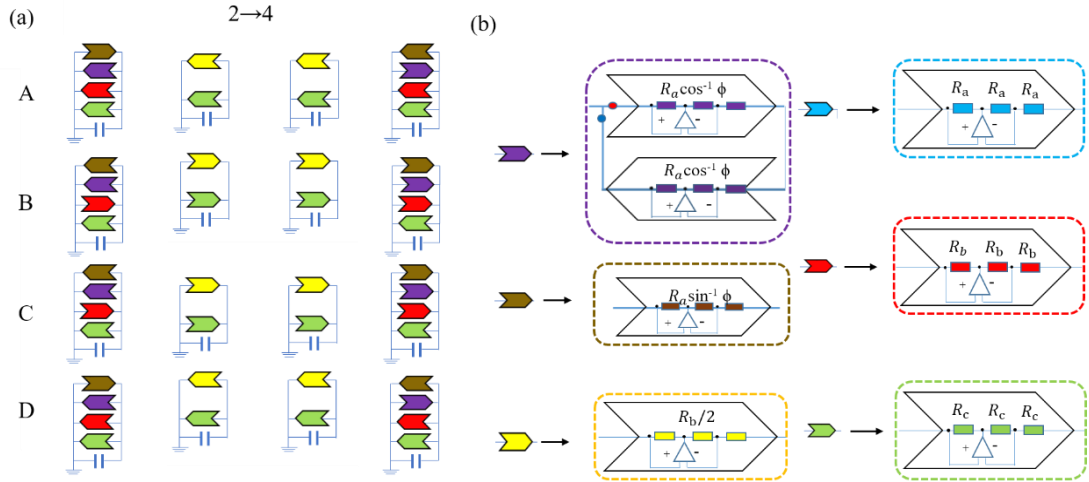

Figure S5 (a) The groundings of the four unit cells from 2 to 4 in the T junction. (b) The corresponding structures of different INICs.

**Supplementary Note 6. The photograph of the fabricated T junction in the electric circuit**

Fig. S6 exhibits the fabricated circuit for the T junction. 1-4 correspond to the four positions in the T junction.

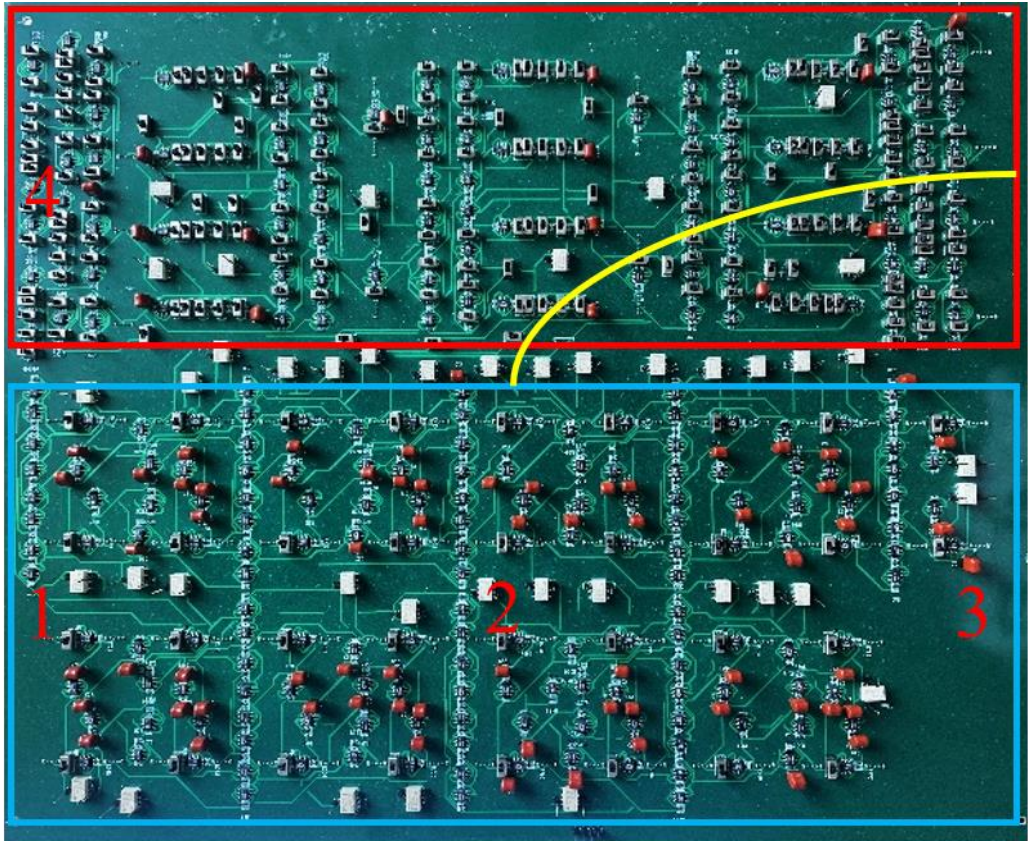

Figure S6 The photograph of the T junction in the experiment.

**Supplementary Note 7. The energy eigenvalue with perturbation during the braiding process**

Fig. S7 shows the energy eigenvalue of the circuit T junction in eight evolution steps with perturbation. We construct 10% coupling randomness. The energies of the edge states (in red) are found to slightly deviate from zero by randomness. But it is still well separated from the bulk spectrum, which indicates that the braiding process is robust against perturbation.

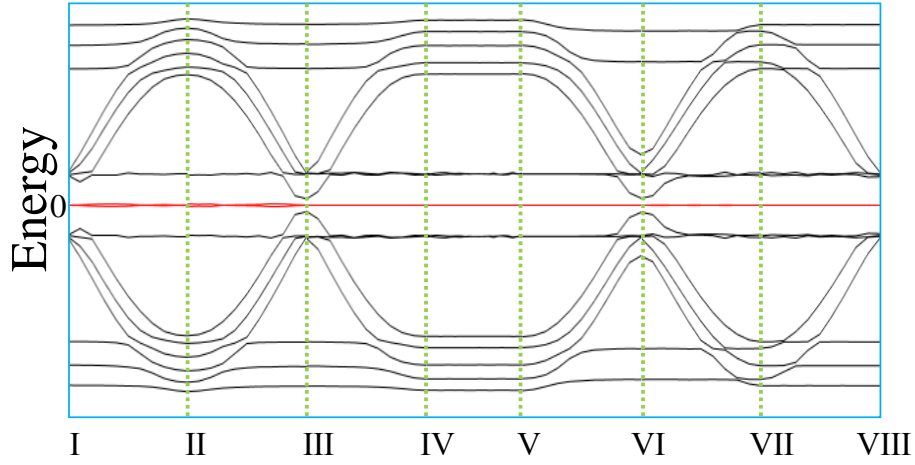

Figure S7 The energy eigenvalue with perturbation during the braiding process.

**Supplementary Note 8. Braiding matrix under the framework of quantum theory**

In this section, we describe the braiding matrix in quantum theory. The exchange operation $U_{\gamma'_i \gamma'_j}$ of the zero modes ($\gamma'_i$ and $\gamma'_j$) in braiding can be expressed as

$$U_{\gamma'_i \gamma'_j} : \begin{cases} \gamma'_i \to -\gamma'_j \\ \gamma'_j \to \gamma'_i \end{cases}. \tag{S5}$$

If we normalize the zero modes by $\{\gamma'_i, \gamma'_j\} = 2\delta_{ij}$, then the expression for $U_{\gamma'_i \gamma'_j}$ is $U_{\gamma'_i \gamma'_j} = \frac{1}{\sqrt{2}}(1 + \gamma'_i \gamma'_j)$. We have $U_{\gamma'_i \gamma'_j} \gamma'_i U_{\gamma'_i \gamma'_j}^{-1} = -\gamma'_j$ and $U_{\gamma'_i \gamma'_j} \gamma'_j U_{\gamma'_i \gamma'_j}^{-1} = \gamma'_i$. The two zero modes can be combined into a single complex mode as $c_a = \frac{1}{2}(\gamma'_1 + i\gamma'_2)$ and $c_a^\dagger = \frac{1}{2}(\gamma'_1 - i\gamma'_2)$. Under the representation of $c_a$, the exchange operator can also be written as

$$U_{\gamma'_i \gamma'_j} = \exp\left[i\frac{\pi}{4}(2c_a^\dagger c_a - 1)\right] = \exp\left(i\frac{\pi}{4}\sigma_z\right) = \begin{pmatrix} e^{-i\pi/4} & 0 \\ 0 & e^{i\pi/4} \end{pmatrix}. \tag{S4}$$

Based on $|0\rangle$ and $|1\rangle$, we have

$$U_{\gamma'_i \gamma'_j} \begin{pmatrix} |0\rangle \\ |1\rangle \end{pmatrix} = \begin{pmatrix} e^{-i\pi/4} & 0 \\ 0 & e^{i\pi/4} \end{pmatrix} \begin{pmatrix} |0\rangle \\ |1\rangle \end{pmatrix}. \tag{S5}$$

It indicates that the braiding process acts as a $\pi/2$ phase gate.

**Supplementary Note 9. The braiding matrix in circuit**

In fact, the corresponding braiding matrix can also be obtained in the circuit. $V_1(0)$ and $V_3(0)$ are the voltages of two edge states ($\gamma_i$ and $\gamma_j$), which correspond to Majorana-like zero modes. After eight steps, the braiding process is achieved with the result

$$\begin{aligned} V_1(0) &\to V_3(0) \\ V_3(0) &\to -V_1(0) \end{aligned}. \quad (S6)$$

We can write it as the form $i\sigma_y \begin{pmatrix} V_1(0) \\ V_3(0) \end{pmatrix} = \begin{pmatrix} V_3(0) \\ -V_1(0) \end{pmatrix}$. If we express this derivation in the form of the braiding in quantum theory, it also satisfies the similar braiding relation connected by the braiding operation $U_{\gamma_i \gamma_j}$. Here, when we regard the edge state $V_1(0)$ and $V_3(0)$ as two Majorana-like zero modes, and consider the braiding operation $U_{\gamma_i \gamma_j}$ is applied to the left and right sides of modes, we can have the similar relation as in the braiding in quantum theory,

$$\begin{aligned} \left[ U_{AB} \begin{pmatrix} V_1 \\ V_3 \end{pmatrix} \right]^T U_{AB}^{-1} &= \left[ \exp\left[\frac{i\pi}{4}\sigma_y\right] \begin{pmatrix} V_1 \\ V_3 \end{pmatrix} \right]^T \exp\left[-\frac{i\pi}{4}\sigma_y\right] \\ &= \frac{1}{2}\left[\begin{pmatrix} 1 & 1 \\ -1 & 1 \end{pmatrix}\begin{pmatrix} V_1 \\ V_3 \end{pmatrix}\right]^T \begin{pmatrix} 1 & -1 \\ 1 & 1 \end{pmatrix} = \frac{1}{2}(V_1+V_3, -V_1+V_3)\begin{pmatrix} 1 & -1 \\ 1 & 1 \end{pmatrix} = (V_3, -V_1) \end{aligned} \quad (S7)$$

Here, the transformation $U_{AB}$ has the form $U_{AB} = \exp\left[\frac{i\pi}{4}\sigma_y\right]$, and it is the braiding matrix for the zero modes in the matrix representation. It is necessary to know how the braiding acts on the initial state $|0)$ and $|1)$, which correspond to $|0\rangle$ and $|1\rangle$ in the qubit representation. We construct the initial state $|0)$ and $|1)$ to be

$$\begin{pmatrix} |0) \\ |1) \end{pmatrix} = U_{basis} \begin{pmatrix} V_1 \\ V_3 \end{pmatrix}, \quad (S8)$$

where $U_{basis} = \frac{1}{\sqrt{2}}\begin{pmatrix} 1 & i \\ 1 & -i \end{pmatrix}$. We can associate $V_1$ and $V_3$ to the operators $\gamma_i$ and $\gamma_j$. So, the operators $f$ and $f^\dagger$ can be expressed as $f = (\gamma_i + i\gamma_j)/2$ and $f^\dagger = (\gamma_i - i\gamma_j)/2$. The states $|0)$ and $|1)$ are defined as empty and the occupied states with respect to operator $f$,

$$f|0\rangle = 0, \ f^\dagger|1\rangle = 0, \ f^\dagger|0\rangle = |1\rangle, \ f|1\rangle = |0\rangle. \quad (S9)$$

The braiding of two edge states can be expressed as

$$U_{\gamma_i\gamma_j} = \frac{1}{\sqrt{2}}\left[1 + i\left(f^\dagger - f\right)\left(f_1^\dagger + f\right)\right]. \tag{S10}$$

By operating this to state $|0\rangle$ and $|1\rangle$, we find

$$U_{\gamma_i\gamma_j}\begin{pmatrix}|0\rangle \\ |1\rangle\end{pmatrix} = \begin{pmatrix} e^{-i\pi/4} & 0 \\ 0 & e^{i\pi/4} \end{pmatrix}\begin{pmatrix}|0\rangle \\ |1\rangle\end{pmatrix}. \tag{S11}$$

$U_{\gamma_i\gamma_j}$ indicates the braiding of two edge states $\gamma_i$ and $\gamma_j$. It corresponds exactly to the braiding matrix in quantum theory. Moreover, the braiding matrix in the circuit $U_{\gamma_i\gamma_j}$ acting on initial states $|0\rangle$ and $|1\rangle$ can also be connected to the braiding process between voltages of two edge states as $U_{\gamma_i\gamma_j}\begin{pmatrix}|0\rangle \\ |1\rangle\end{pmatrix} = U_{basis}U_{AB}U_{basis}^{-1}U_{basis}\begin{pmatrix}V_1 \\ V_3\end{pmatrix} = \begin{pmatrix} e^{-i\pi/4} & 0 \\ 0 & e^{i\pi/4} \end{pmatrix}\begin{pmatrix}|0\rangle \\ |1\rangle\end{pmatrix}.$

To experimentally test the theoretical analysis, we fabricate an electric circuit as shown in Fig. S6. We use the DC signal source (UTP1306S) to create the required DC voltage signal. The digital storage oscilloscope (Agilent Technologies Infiniivision DSO7104B) is used to measure the voltage signal at each node simultaneously. After the whole braiding process from I to IX, we measure the voltages at different nodes in 1 and 3, which reads

$$\begin{pmatrix}V_A^1 \\ V_B^1 \\ V_C^1 \\ V_D^1\end{pmatrix} = \begin{pmatrix}6.08 \\ 6.07 \\ -0.4 \\ -0.42\end{pmatrix} \text{ and } \begin{pmatrix}V_A^3 \\ V_B^3 \\ V_C^3 \\ V_D^3\end{pmatrix} = \begin{pmatrix}0.23 \\ -0.33 \\ -6.71 \\ 6.67\end{pmatrix}. \tag{S12}$$

$V_j^i$ is the voltage in the node $j$ on the position $i$ of T junction. After normalization, we get

$$\begin{pmatrix}V_A^1 \\ V_B^1 \\ V_C^1 \\ V_D^1\end{pmatrix} = \begin{pmatrix}0.7061 \\ 0.7049 \\ -0.0465 \\ -0.0488\end{pmatrix} \text{ and } \begin{pmatrix}V_A^3 \\ V_B^3 \\ V_C^3 \\ V_D^3\end{pmatrix} = \begin{pmatrix}0.0243 \\ -0.0348 \\ -0.7086 \\ 0.7044\end{pmatrix}. \tag{S13}$$

We write it into the form

$$V_A^1 + iV_C^1 = 0.7061 - 0.0465i,$$

$$V_B^1 + iV_D^1 = 0.7049 - 0.0488i,$$

$$V_A^3 + iV_C^3 = 0.0243 - 0.7086i,$$

$$V_B^3 + iV_D^3 = -0.0348 + 0.7044i. \tag{S14}$$

We project it on the initial state $\frac{1}{\sqrt{2}}(1,1)$ and $\frac{1}{\sqrt{2}}(-i,i)$, and write it as the form

$$\begin{bmatrix} \begin{pmatrix} V_A^1 + iV_C^1 \\ V_B^1 + iV_D^1 \end{pmatrix} \cdot \frac{1}{\sqrt{2}}(-i \quad i) & \begin{pmatrix} V_a^1 + iV_c^1 \\ V_b^1 + iV_d^1 \end{pmatrix} \cdot \frac{1}{\sqrt{2}}(1 \quad 1) \\ \begin{pmatrix} V_A^3 + iV_C^3 \\ V_B^3 + iV_D^3 \end{pmatrix} \cdot \frac{1}{\sqrt{2}}(-i \quad i) & \begin{pmatrix} V_A^3 + iV_C^3 \\ V_B^3 + iV_D^3 \end{pmatrix} \cdot \frac{1}{\sqrt{2}}(1 \quad 1) \end{bmatrix} = \begin{bmatrix} 0.0016 - 0.0008i & 0.9977 - 0.0673i \\ -0.9991 - 0.0418i & -0.0075 - 0.003i \end{bmatrix},$$

(S15)

which corresponds to $i\sigma_y$. These equations describe the braiding behavior of Majorana-like zero-energy mode. $U_{AB}$ is related to the braiding of two zero-energy mode $i\sigma_y$,

$$U_{AB} = \begin{bmatrix} 0.7088 + 0.0068i & 0.704 - 0.0466i \\ -0.7048 - 0.0304i & 0.7024 + 0.0053i \end{bmatrix}. \tag{S16}$$

The braiding matrix $U_{\gamma_i\gamma_j}$ acts on the one-qubit state as

$$U_{\gamma_i\gamma_j}\begin{pmatrix} \vec{\psi}_{|0\rangle} \\ \vec{\psi}_{|1\rangle} \end{pmatrix} = U_{basis} U_{AB} U_{basis} \begin{pmatrix} \vec{\psi}_A \\ \vec{\psi}_B \end{pmatrix} = \begin{pmatrix} 0.7 - 0.7i & 0.04 \\ -0.04 & 0.71 + 0.71i \end{pmatrix} \begin{pmatrix} \vec{\psi}_{|0\rangle} \\ \vec{\psi}_{|1\rangle} \end{pmatrix}. \tag{S17}$$

The result of $U_{\gamma_i\gamma_j}$ is shown in Fig. 2(e) of the main text.

**Supplementary Note 10. The theoretical description and experimental data for H gate**

According to the theory of topological quantum computing, four of Majorana-like zero modes $\gamma_1'$, $\gamma_2'$, $\gamma_3'$ and $\gamma_4'$ are required to construct the H gate. We choose $\gamma_1'$ and $\gamma_2'$ as the basis. Based on the above braiding process $U_{\gamma_i'\gamma_j'}$, the exchange operations for these four zero modes can be expressed as

$$U_{\gamma_1'\gamma_2'} = U_{\gamma_3'\gamma_4'} = e^{-i\pi/4}\begin{bmatrix} 1 & 0 \\ 0 & i \end{bmatrix}, \quad U_{\gamma_2'\gamma_3'} = \frac{1}{\sqrt{2}}\begin{bmatrix} 1 & -i \\ -i & 1 \end{bmatrix}. \tag{S18}$$

H gate can be implemented by the following exchange operations,

$$U_{\gamma_1'\gamma_2',1} U_{\gamma_2'\gamma_3'} U_{\gamma_1'\gamma_2',2} = e^{i\pi/2}\frac{1}{\sqrt{2}}\begin{bmatrix} 1 & 1 \\ 1 & -1 \end{bmatrix} \sim H, \tag{S19}$$

where $U_{i,j}$ is the $j$th exchange operations $U_i$. It can also be simulated by designing the electric circuit. Two pairs (four) of edge states corresponding to Majorana-like zero modes $\gamma_1$, $\gamma_2$, $\gamma_3$ and $\gamma_4$ are used. Based on the braiding matrix in the circuit of Supplementary Note

6, the exchange operations for these four zero modes can be expressed as $U_{\gamma_1\gamma_2} = e^{-i\pi/4}\begin{bmatrix} 1 & 0 \\ 0 & i \end{bmatrix}$,

$U_{\gamma_2\gamma_3} = \frac{1}{\sqrt{2}}\begin{bmatrix} 1 & -i \\ -i & 1 \end{bmatrix}$. To achieve the H gate in the circuit, the braiding of these edge states also needs three exchange operations using T junctions $U_{\gamma_1\gamma_2,1}U_{\gamma_2\gamma_3}U_{\gamma_1\gamma_2,2} = e^{i\pi/2}\frac{1}{\sqrt{2}}\begin{bmatrix} 1 & 1 \\ 1 & -1 \end{bmatrix} \sim H$. It corresponds exactly to the H gate matrix in quantum theory.

Furthermore, we fabricate the corresponding circuit network to measure the matrix of output voltages for the H gate in the experiment. The measured exchange operations are

$$U_{\gamma_1\gamma_2,1} = \begin{bmatrix} 0.7-0.7i & 0.04 \\ -0.04 & 0.71+0.71i \end{bmatrix},$$

$$U_{\gamma_2\gamma_3} = \frac{1}{\sqrt{2}}\begin{bmatrix} 0.98 & -0.11-0.99i \\ 0.11-0.99i & 0.98 \end{bmatrix},$$

$$U_{\gamma_1\gamma_2,2} = \begin{bmatrix} 0.78-0.78i & 0 \\ 0.01 & 0.65+0.65i \end{bmatrix}. \tag{S20}$$

So, the measured matrix for the H gate is

$$H = U_{\gamma_1\gamma_2,1}U_{\gamma_2\gamma_3}U_{\gamma_1\gamma_2,2} = e^{-i\pi/2}\frac{1}{\sqrt{2}}\begin{bmatrix} 1.11-0.03i & 0.88-0.07i \\ 1.06+0.1i & -0.93-0.02i \end{bmatrix}. \tag{S21}$$

**Supplementary Note 11. The theoretical description and experimental data for Z gate**

Based on the above exchange operations for the four zero modes in Eq. (S18), the corresponding Z gate can be implemented by the following switching operations,

$$U_{\gamma_1\gamma_2',1}U_{\gamma_1\gamma_2,2} = e^{-i\pi/2}\begin{bmatrix} 1 & 0 \\ 0 & -1 \end{bmatrix} \sim Z. \tag{S22}$$

It can also be simulated by designing the electric circuit. The braiding of the Z gate in the circuit can be expressed as

$$U_{\gamma_1\gamma_2,1}U_{\gamma_1\gamma_2,2} = e^{-i\pi/2}\begin{bmatrix} 1 & 0 \\ 0 & -1 \end{bmatrix} \sim Z. \tag{S23}$$

It corresponds exactly to the Z gate matrix in quantum theory.

Furthermore, we also measure the matrix for the Z gate in the experiment. The measured

exchange operations are

$$U_{\gamma_1\gamma_2,1} = \begin{bmatrix} 0.7-0.7i & 0.04 \\ -0.04 & 0.71+0.71i \end{bmatrix},$$

$$U_{\gamma_1\gamma_2,2} = \begin{bmatrix} 0.77-0.77i & 0.01 \\ 0 & 0.69+0.68i \end{bmatrix}. \tag{S24}$$

So, the measured matrix for the Z gate is

$$Z = U_{\gamma_1\gamma_2,1}U_{\gamma_1\gamma_2,2} = e^{-i\pi/2}\begin{bmatrix} 1.08 & -0.02+0.03i \\ -0.03-0.03i & -0.97+0.01i \end{bmatrix}. \tag{S25}$$

**Supplementary Note 12. The theoretical description and experimental data for X gate**

The X gate can be implemented by the following switching operations,

$$U_{\gamma'_2\gamma'_3,1}U_{\gamma'_2\gamma'_3,2} = e^{-i\pi/2}\begin{bmatrix} 0 & 1 \\ 1 & 0 \end{bmatrix} \sim X. \tag{S26}$$

The braiding of X gate in the circuit network can be expressed as

$$U_{\gamma_2\gamma_3,1}U_{\gamma_2\gamma_3,2} = e^{-i\pi/2}\begin{bmatrix} 0 & 1 \\ 1 & 0 \end{bmatrix} \sim X. \tag{S27}$$

It corresponds exactly to the X gate matrix in quantum theory.

We also measure the matrix of output voltages for the X gate in the experiment. The measured exchange operations are

$$U_{\gamma_2\gamma_3,1} = \frac{1}{\sqrt{2}}\begin{bmatrix} 0.99+0.01i & -0.06-i \\ 0.04-i & 1+0.01i \end{bmatrix},$$

$$U_{\gamma_2\gamma_3,2} = \frac{1}{\sqrt{2}}\begin{bmatrix} 1.01-0.09i & 0.1-i \\ 0.08-i & 1-0.09i \end{bmatrix}. \tag{S28}$$

So, the measured matrix for the X gate is

$$X = U_{\gamma_2\gamma_3,1}U_{\gamma_2\gamma_3,2} = e^{-i\pi/2}\begin{bmatrix} 0.05 & 0.99-0.02i \\ 1+0.02i & 0.11 \end{bmatrix}. \tag{S29}$$

From the experimental results of the measured matrix for the X gate, as shown in Fig. S8, it is seen clearly that the input states for $|0\rangle$ and $|1\rangle$ are successfully transformed into $|1\rangle$ and $|0\rangle$, respectively.

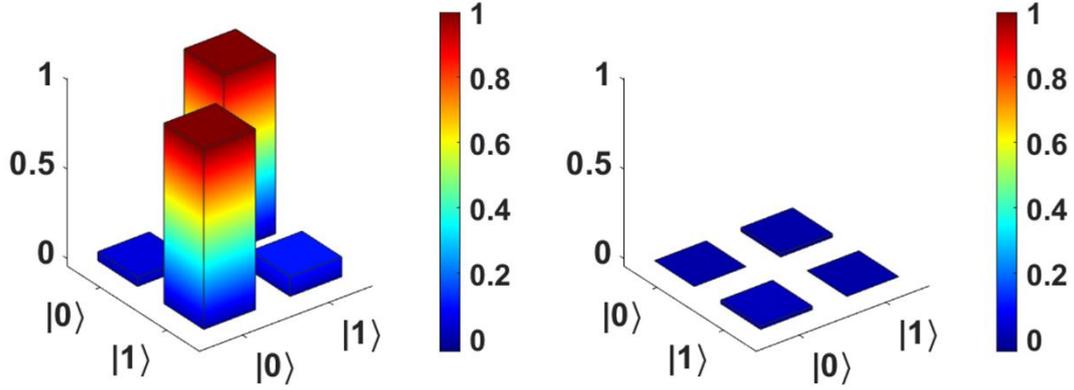

Figure S8 The output voltage matrices of X gate.

## Supplementary Note 13. The theoretical description and experimental data for CNOT gate

According to the theory of topological quantum computing, six Majorana-like zero modes $\gamma'_1$, $\gamma'_2$, $\gamma'_3$, $\gamma'_4$, $\gamma'_5$ and $\gamma'_6$ are required to construct a CNOT gate. We choose $\gamma'_1$, $\gamma'_2$, $\gamma'_5$ and $\gamma'_6$ as the basis. The exchange operations for these six zero modes can be expressed as

$$U_{\gamma'_1\gamma'_2}=e^{-i\pi/4}diag(1,1,i,i), U_{\gamma'_3\gamma'_4}=e^{-i\pi/4}diag(1,i,i,1), U_{\gamma'_5\gamma'_6}=e^{-i\pi/4}diag(1,i,1,i),$$

$$U_{\gamma'_2\gamma'_3}=\frac{1}{\sqrt{2}}\begin{bmatrix} 1 & 0 & -i & 0 \\ 0 & 1 & 0 & -i \\ -i & 0 & 1 & 0 \\ 0 & -i & 0 & 1 \end{bmatrix}, U_{\gamma'_4\gamma'_5}=\frac{1}{\sqrt{2}}\begin{bmatrix} 1 & -i & 0 & 0 \\ -i & 1 & 0 & 0 \\ 0 & 0 & 1 & -i \\ 0 & 0 & -i & 1 \end{bmatrix}. \tag{S30}$$

The CNOT gate can be implemented by the following switching operations,

$$U_{\gamma'_3\gamma'_4,1}U_{\gamma'_5\gamma'_4,1}U_{\gamma'_4\gamma'_3}U_{\gamma'_2\gamma'_1}U_{\gamma'_6\gamma'_5}U_{\gamma'_5\gamma'_4,2}U_{\gamma'_3\gamma'_4,2}=e^{-i\pi/4}\begin{bmatrix} 1 & 0 & 0 & 0 \\ 0 & 1 & 0 & 0 \\ 0 & 0 & 0 & 1 \\ 0 & 0 & 1 & 0 \end{bmatrix} \sim CNOT. \tag{S31}$$

Such a CNOT gate can also be simulated by designing the electric circuit. Six edge states corresponding to Majorana-like zero modes $\gamma_1$, $\gamma_2$, $\gamma_3$, $\gamma_4$, $\gamma_5$ and $\gamma_6$ are used. The exchange operations for these four zero modes can be expressed as

$$U_{\gamma_1\gamma_2}=e^{-i\pi/4}diag(1,1,i,i), U_{\gamma_3\gamma_4}=e^{-i\pi/4}diag(1,i,i,1), U_{\gamma_5\gamma_6}=e^{-i\pi/4}diag(1,i,1,i),$$

$$U_{\gamma_2\gamma_3} = \frac{1}{\sqrt{2}}\begin{bmatrix} 1 & 0 & -i & 0 \\ 0 & 1 & 0 & -i \\ -i & 0 & 1 & 0 \\ 0 & -i & 0 & 1 \end{bmatrix}, \quad U_{\gamma_4\gamma_5} = \frac{1}{\sqrt{2}}\begin{bmatrix} 1 & -i & 0 & 0 \\ -i & 1 & 0 & 0 \\ 0 & 0 & 1 & -i \\ 0 & 0 & -i & 1 \end{bmatrix}. \quad (S32)$$

The corresponding CNOT gate in the circuit can be expressed as,

$$U_{\gamma_3\gamma_4,1}U_{\gamma_5\gamma_4,1}U_{\gamma_4\gamma_3}U_{\gamma_2\gamma_1}U_{\gamma_6\gamma_5}U_{\gamma_5\gamma_4,2}U_{\gamma_3\gamma_4,2} = e^{-i\pi/4}\begin{bmatrix} 1 & 0 & 0 & 0 \\ 0 & 1 & 0 & 0 \\ 0 & 0 & 0 & 1 \\ 0 & 0 & 1 & 0 \end{bmatrix} \sim CNOT. \quad (S33)$$

It corresponds exactly to the CNOT gate matrix in quantum theory.

Furthermore, we fabricate the corresponding circuit network to measure the matrix for the CNOT gate in the experiment. The measured exchange operations are

$$U_{\gamma_3\gamma_4,1} = \begin{bmatrix} 0.7-0.7i & 0 & 0.04 & 0 \\ 0 & 0.71+0.71i & 0 & -0.04 \\ -0.04 & 0 & 0.71+0.71i & 0 \\ 0 & 0.04 & 0 & 0.7-0.7i \end{bmatrix},$$

$$U_{\gamma_5\gamma_4,1} = \frac{1}{\sqrt{2}}\begin{bmatrix} 0.98 & -0.11-0.99i & 0 & 0 \\ 0.11-0.99i & 0.98 & 0 & 0. \\ 0 & 0 & 0.98 & -0.11-0.99i \\ 0 & 0 & 0.11-0.99i & 0.98 \end{bmatrix},$$

$$U_{\gamma_4\gamma_3,1} = \begin{bmatrix} 0.75-0.75i & 0 & 0 & 0 \\ 0 & 0.65+0.65i & 0 & 0.01 \\ 0.01 & 0 & 0.65+0.65i & 0 \\ 0 & 0 & 0 & 0.75-0.75i \end{bmatrix},$$

$$U_{\gamma_2\gamma_1} = \begin{bmatrix} 0.7-0.7i & 0 & 0 & 0.04 \\ 0 & 0.7-0.7i & 0.04 & 0 \\ 0 & -0.04 & 0.71+0.71i & 0 \\ -0.04 & 0 & 0 & 0.71+0.71i \end{bmatrix},$$

$$U_{\gamma_6\gamma_5} = \begin{bmatrix} 0.75-0.75i & -0.02 & 0 & 0 \\ 0.03 & 0.65+0.65i & 0 & 0 \\ 0 & 0 & 0.75-0.75i & -0.02 \\ 0 & 0 & 0.03 & 0.65+0.65i \end{bmatrix},$$

$$U_{\gamma_5\gamma_4,2} = \frac{1}{\sqrt{2}} \begin{bmatrix} 1+0.04i & -0.07-i & 0 & 0 \\ -0.04-i & 1+0.05i & 0 & 0. \\ 0 & 0 & 1+0.04i & -0.07-i \\ 0 & 0 & -0.04-i & 1+0.05i \end{bmatrix},$$

$$U_{\gamma_3\gamma_4,2} = \begin{bmatrix} 0.72-0.72i & -0.08 & 0 & 0 \\ 0.05 & 0.7+0.7i & 0 & 0 \\ 0 & 0 & 0.72-0.72i & -0.08 \\ 0 & 0 & 0.05 & 0.7+0.7i \end{bmatrix}. \quad \text{(S34)}$$

So, the measured matrix for the CNOT gate is

$$CNOT = U_{\gamma_3\gamma_4,1}U_{\gamma_5\gamma_4,1}U_{\gamma_4\gamma_3}U_{\gamma_2\gamma_1}U_{\gamma_6\gamma_5}U_{\gamma_5\gamma_4,2}U_{\gamma_3\gamma_4,2} = e^{-i\pi/4}\begin{bmatrix} 0.97-0.02i & -0.17-0.14i & -0.03-0.03i & 0.03-0.03i \\ -0.13-0.05i & 0.96+0.06i & -0.03-0.02i & -0.03-0.03i \\ -0.03-0.02i & 0.02-0.02i & -0.05-0.05i & 0.98-0.08i \\ 0.03-0.03i & 0.03-0.03i & 0.95+0.08i & 0.02+0.06i \end{bmatrix}. \quad \text{(S35)}$$

**Supplementary Note 14. The theoretical description and experimental data for CZ gate**

The CZ gate can be implemented by the following switching operations,

$$U_{\gamma_3\gamma_4'}U_{\gamma_2\gamma_1'}U_{\gamma_6\gamma_5'} = e^{-i\pi/4}\begin{bmatrix} 1 & 0 & 0 & 0 \\ 0 & 1 & 0 & 0 \\ 0 & 0 & 1 & 0 \\ 0 & 0 & 0 & -1 \end{bmatrix} \sim CZ. \quad \text{(S36)}$$

Such a CZ gate can also be simulated by designing the electric circuit. The corresponding CZ gate in the circuit can be expressed as,

$$U_{\gamma_3\gamma_4}U_{\gamma_2\gamma_1}U_{\gamma_6\gamma_5} = e^{-i\pi/4}\begin{bmatrix} 1 & 0 & 0 & 0 \\ 0 & 1 & 0 & 0 \\ 0 & 0 & 1 & 0 \\ 0 & 0 & 0 & -1 \end{bmatrix} \sim CZ. \quad \text{(S37)}$$

It corresponds exactly to the CZ gate matrix in quantum theory.

Furthermore, we fabricate the corresponding circuit network to measure the matrix for the CZ gate in the experiment. The measured exchange operations are

$$U_{\gamma_3\gamma_4} = \begin{bmatrix} 0.7-0.7i & 0 & 0.04 & 0 \\ 0 & 0.71+0.71i & 0 & -0.04 \\ -0.04 & 0 & 0.71+0.71i & 0 \\ 0 & 0.04 & 0 & 0.7-0.7i \end{bmatrix},$$

$$U_{\gamma_1\gamma_2} = \begin{bmatrix} 0.7-0.7i & 0 & 0 & 0.04 \\ 0 & 0.7-0.7i & 0.04 & 0 \\ 0 & -0.04 & 0.71+0.71i & 0 \\ -0.04 & 0 & 0 & 0.71+0.71i \end{bmatrix},$$

$$U_{\gamma_5\gamma_6} = \begin{bmatrix} 0.7-0.7i & 0.04 & 0 & 0 \\ -0.04 & 0.71+0.71i & 0 & 0 \\ 0 & 0 & 0.7-0.7i & 0.04 \\ 0 & 0 & -0.04 & 0.71+0.71i \end{bmatrix}. \quad (S38)$$

So, the measured matrix in the experiment for the CZ gate is

$$CZ = U_{\gamma_3\gamma_4} U_{\gamma_2\gamma_1} U_{\gamma_5\gamma_6} = e^{-i\pi/4} \begin{bmatrix} 0.97 & -0.03+0.03i & 0.03-0.03i & 0.03+0.03i \\ 0.03+0.03i & 1 & -0.03-0.03i & 0.03+0.03i \\ -0.03-0.03i & 0.03-0.03i & 1 & -0.03+0.03i \\ 0.03-0.03i & 0.03-0.03i & -0.03-0.03i & -1 \end{bmatrix}. \quad (S39)$$

**Supplementary Note 15. Theoretical scheme for Grover's search**

The topological quantum computation schemes for the other cases ($00$, $\pi 0$ and $\pi\pi$) of Grover's search are exhibited in Fig. S9.

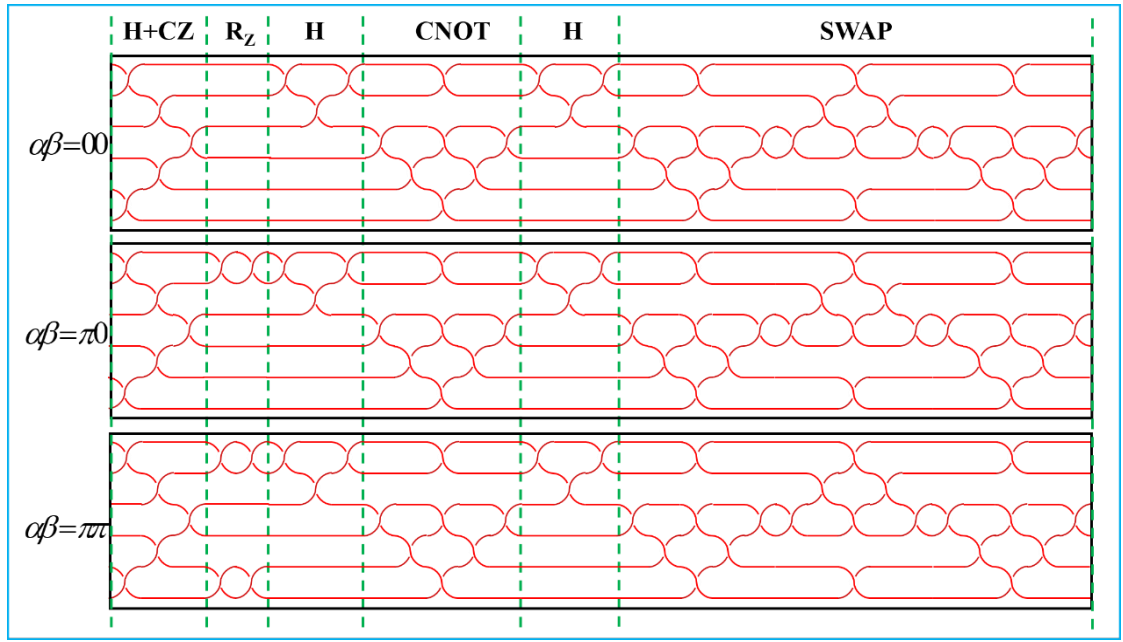

Figure S9 The quantum theoretical schemes for Grover's search with different $\alpha\beta$.

These topological quantum computation schemes correspond to the quantum route diagrams

as shown in Fig. 5(a) one by one. The entire Grover's algorithm $G'_{\alpha\beta}$ can be represented as

$$G'_{\alpha\beta} \sim H_1 H_2 CZ_1 R_z(-\alpha) R_z(-\beta) H_1 CN_1 CN_2 CN_1, \quad \text{(S40)}$$

where $R_z(0) = I$ and $R_z(-\pi) = Z$. So, the matrixes for Grover's search algorithm can be expressed as

$$G'_{00} = U_{\gamma'_1\gamma'_2} U_{\gamma'_2\gamma'_3} U_{\gamma'_5\gamma'_6} U_{\gamma'_4\gamma'_5} U_{\gamma'_3\gamma'_4} I_1 I_2 H_1 CN_1 H_1 CN_1 CN_2 CN_1$$

$$= \begin{bmatrix} 1 & 0 & 0 & 0 \\ 0 & 1 & 0 & 0 \\ 0 & 0 & 1 & 0 \\ 0 & 0 & 0 & -1 \end{bmatrix},$$

$$G'_{01} = U_{\gamma'_1\gamma'_2} U_{\gamma'_2\gamma'_3} U_{\gamma'_5\gamma'_6} U_{\gamma'_4\gamma'_5} U_{\gamma'_3\gamma'_4} I_1 Z_2 H_1 CN_1 H_1 CN_1 CN_2 CN_1$$

$$= \begin{bmatrix} 0 & 1 & 0 & 0 \\ 1 & 0 & 0 & 0 \\ 0 & 0 & 0 & -1 \\ 0 & 0 & 1 & 0 \end{bmatrix},$$

$$G'_{10} = U_{\gamma'_1\gamma'_2} U_{\gamma'_2\gamma'_3} U_{\gamma'_5\gamma'_6} U_{\gamma'_4\gamma'_5} U_{\gamma'_3\gamma'_4} Z_1 I_2 H_1 CN_1 H_1 CN_1 CN_2 CN_1$$

$$= \begin{bmatrix} 0 & 0 & 1 & 0 \\ 0 & 0 & 0 & -1 \\ 1 & 0 & 0 & 0 \\ 0 & 1 & 0 & 0 \end{bmatrix},$$

$$G'_{11} = U_{\gamma'_1\gamma'_2} U_{\gamma'_2\gamma'_3} U_{\gamma'_5\gamma'_6} U_{\gamma'_4\gamma'_5} U_{\gamma'_3\gamma'_4} Z_1 Z_2 H_1 CN_1 H_1 CN_1 CN_2 CN_1$$

$$= \begin{bmatrix} 0 & 0 & 0 & -1 \\ 0 & 0 & 1 & 0 \\ 0 & 1 & 0 & 0 \\ 1 & 0 & 0 & 0 \end{bmatrix}. \quad \text{(S41)}$$

When they act on the initial state $|00\rangle$, the readouts are the marked state,

$$G'_{00}|00\rangle = |00\rangle \ , \ G'_{10}|00\rangle = |10\rangle \ , \ G'_{01}|00\rangle = |01\rangle \ , \ G'_{11}|00\rangle = |11\rangle. \quad \text{(S42)}$$

So, we need only one evolution to identify the hidden marked state in the black box.

**Supplementary Note 16. The photograph of designed and fabricated circuit to achieve Grover's search**

Fig. S10(a) and S10(b) show the abbreviated photograph of the designed and fabricated circuit, respectively. $\gamma_1 - \gamma_6$ represent the six edge states for braiding. In experiment, we construct 7 (4) unit cells in A-C (B-D) of a T junction.

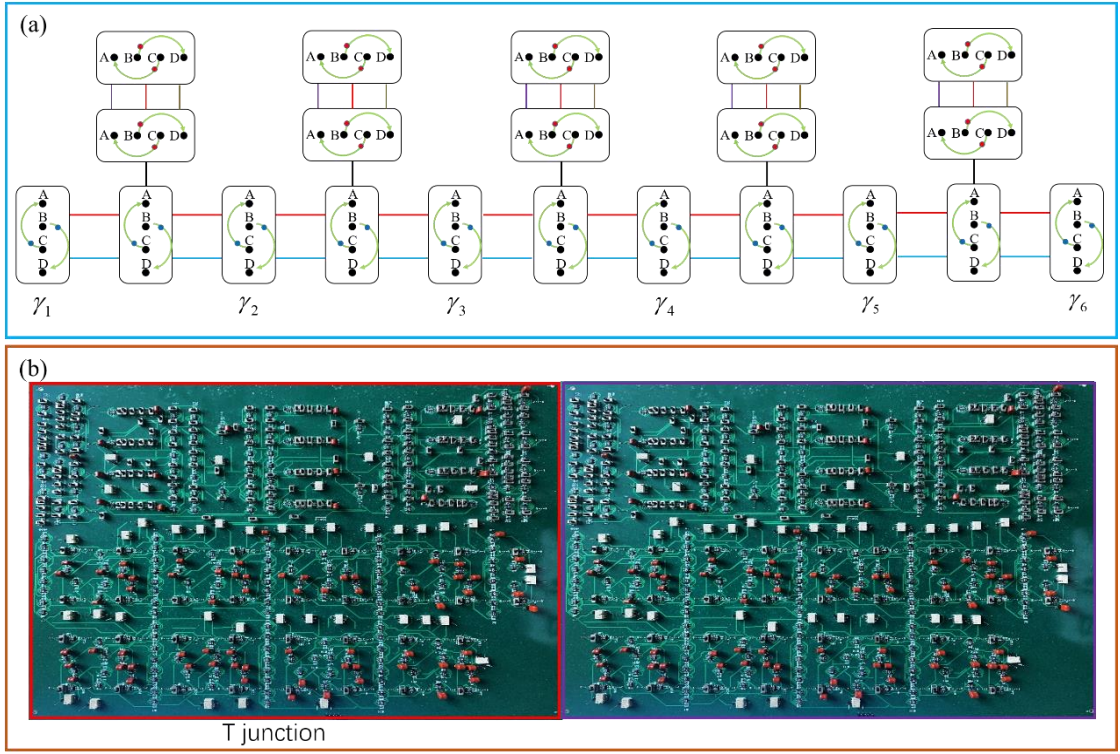

Figure S10 The photograph of the designed and fabricated circuit to achieve Grover's search. (a) The abbreviated designed circuit. There are more units in the experiment. (b) Part of the fabricated circuit in the experiment, which composed of several T junctions.

**Supplementary Note 17. Circuit theory and experimental dates for Grover's search**

Based on the above theoretical descriptions of the correspondence between quantum scheme and circuit theory in two-qubit gate operations, we also use six edge states to prepare the two-qubit input $|00)$ which corresponds to $|00\rangle$ in the quantum scheme. According to Eq. (S40), the braiding for Grover's search algorithm in the circuit $G_{\alpha\beta}$ can be expressed as

$$G_{00} = U_{\gamma_1\gamma_2} U_{\gamma_2\gamma_3} U_{\gamma_5\gamma_6} U_{\gamma_4\gamma_5} U_{\gamma_3\gamma_4} I_1 I_2 H_1 CN_1 H_1 CN_1 CN_2 CN_1,$$

$$G_{01} = U_{\gamma_1\gamma_2} U_{\gamma_2\gamma_3} U_{\gamma_5\gamma_6} U_{\gamma_4\gamma_5} U_{\gamma_3\gamma_4} I_1 Z_2 H_1 CN_1 H_1 CN_1 CN_2 CN_1$$

$$G_{10} = U_{\gamma_1\gamma_2} U_{\gamma_2\gamma_3} U_{\gamma_5\gamma_6} U_{\gamma_4\gamma_5} U_{\gamma_3\gamma_4} Z_1 I_2 H_1 CN_1 H_1 CN_1 CN_2 CN_1,$$

$$G_{11} = U_{\gamma_1\gamma_2} U_{\gamma_2\gamma_3} U_{\gamma_5\gamma_6} U_{\gamma_4\gamma_5} U_{\gamma_3\gamma_4} Z_1 Z_2 H_1 CN_1 H_1 CN_1 CN_2 CN_1. \quad \textbf{(S43)}$$

We fabricate the corresponding circuit network to measure the matrix in the experiment. The

results are

$$G_{00} = \begin{bmatrix} 4.14-0.14i & -0.05-1.42i & -1.26-1.37i & -0.95-0.58i \\ -0.84-0.35i & 3.21-0.03i & 0.74-1.18i & -0.93-0.1i \\ -1.02-0.66i & -0.26+0.26i & 4.45+1.17i & 0.08+0.74i \\ -0.03+0.51i & 0.1-0.15i & 0.07+0.11i & -3.38-0.21i \end{bmatrix},$$

$$G_{10} = \begin{bmatrix} -0.74-0.54i & 0.25i & 4.46+0.46i & -0.03+0.7i \\ 0.21+0.38i & 0.33-0.21i & -0.11-0.15i & -3.47+0.12i \\ 4.14+0.22i & -1.33i & -0.76-1.41i & -0.64-0.53i \\ -0.58-0.31i & 3.26+0.24i & 0.6-1.1i & -1.21-0.28i \end{bmatrix},$$

$$G_{01} = \begin{bmatrix} -0.45-0.79i & 3.2-0.12i & 0.43-1.24i & -1.11 \\ 4.42-0.26i & -0.15-1.18i & -1.23-1.22i & -1.06-0.68i \\ -0.34+0.6i & 0.14-0.2i & 0.61-0.22i & -3.42-0.16i \\ -1.17-0.57i & -0.12+0.3i & 4.75+1.15i & 0.2+0.41i \end{bmatrix},$$

$$G_{11} = \begin{bmatrix} -0.03+0.68i & 0.42-0.25i & 0.68-0.6i & -3.48+0.13i \\ -0.84-0.45i & 0.07+0.11i & 4.64+0.38i & 0.23+0.4i \\ -0.42-0.63i & 3.28+0.22i & 0.38-1.43i & -1.35-0.16i \\ 4.36+0.11i & 0.09-1.09i & -0.73-1.3i & -0.82-0.88i \end{bmatrix}. \quad \textbf{(S44)}$$

When they act on the initial state $|00\rangle$ after normalization, the output is,

$$G_{00}|00\rangle = \sqrt{0.9}|00\rangle + \sqrt{0.04}|01\rangle + \sqrt{0.06}|10\rangle + \sqrt{0}|11\rangle \ ,$$

$$G_{01}|00\rangle = \sqrt{0.01}|00\rangle + \sqrt{0.92}|01\rangle + \sqrt{0}|10\rangle + \sqrt{0.07}|11\rangle \ ,$$

$$G_{10}|00\rangle = \sqrt{0.03}|00\rangle + \sqrt{0}|01\rangle + \sqrt{0.95}|10\rangle + \sqrt{0.02}|11\rangle \ ,$$

$$G_{11}|00\rangle = \sqrt{0}|00\rangle + \sqrt{0.04}|01\rangle + \sqrt{0.01}|10\rangle + \sqrt{0.95}|11\rangle. \quad \textbf{(S45)}$$

It is seen that the ratio of the correct outcome is above 90%. These high-fidelity results indicate the first demonstration of Grover's search algorithm in the circuit.

**Supplementary Note 18. A detailed description of phase rotation**

Steps IV to V and VIII to IX involve phase rotation. For the convenience of the experiment, we replace the variable resistor with several fixed resistors. For example, the variable resistance in purple coupling is $R_a \cos^{-1}\phi$ with $\phi$ from 0 to $\pi$. When $R_a = 1k\Omega$, we use seven resistances $R_a \cos^{-1} 0 = 1K\Omega$, $R_a \cos^{-1}(\pi/6) = 1.15K\Omega$, $R_a \cos^{-1}(\pi/3) = 2K\Omega$,

$R_a \cos^{-1}(\pi/2) = 1M\Omega$, $R_a \cos^{-1}(2\pi/3) = -2K\Omega$, $R_a \cos^{-1}(5\pi/6) = -1.15K\Omega$ and $R_a \cos^{-1}\pi = -1K\Omega$ to replace variable resistance $R_a \cos^{-1}\phi$. Besides, the variable resistance in brown coupling is $R_a \sin^{-1}\phi$ with $\phi$ from 0 to $\pi$. When $R_a = 1k\Omega$, we use seven resistances $R_a \sin^{-1}0 = 1M\Omega$, $R_a \sin^{-1}(\pi/6) = 2K\Omega$, $R_a \sin^{-1}(\pi/3) = 1.15K\Omega$, $R_a \sin^{-1}(\pi/2) = 1K\Omega$, $R_a \sin^{-1}(2\pi/3) = 1.15K\Omega$, $R_a \sin^{-1}(5\pi/6) = 2K\Omega$ and $R_a \sin^{-1}\pi = 1M\Omega$ to replace variable resistance $R_a \sin^{-1}\phi$. In the experiment, we constantly change the resistances and record the voltage signals after each evolution which are used to be the input for the next. Each evolution takes 6ms. Fig. S11(a) and S11(b) show the simulated voltages change with time in step IV to V. We input voltage 10V to node C in 2 and nodes A and B in unit 4. We input voltage -10V to node D in unit 2. The rest of the nodes are connected to the ground. After 42ms, voltages in nodes C and D of unit 2 and A and B of unit 4 become zero. Voltages in nodes A and B of unit 2 and D of unit 4 become 10V. Node C of unit 4 becomes -10V,

$$\begin{pmatrix} V_A^2 = 0 \\ V_B^2 = 0 \\ V_C^2 = 10 \\ V_D^2 = -10 \end{pmatrix}, \begin{pmatrix} V_A^4 = 10 \\ V_B^4 = 10 \\ V_C^4 = 0 \\ V_D^4 = 0 \end{pmatrix} \rightarrow \begin{pmatrix} V_A^2 = 10 \\ V_B^2 = 10 \\ V_C^2 = 0 \\ V_D^2 = 0 \end{pmatrix}, \begin{pmatrix} V_A^4 = 0 \\ V_B^4 = 0 \\ V_C^4 = -10 \\ V_D^4 = 10 \end{pmatrix}. \tag{S46}$$

So, we can get $V_1(0) \rightarrow V_3(0)$ and $V_3(0) \rightarrow -V_1(0)$, which indicates the phase rotation in the step IV to V. Fig. S11(c) and S11(d) show voltages change with time in the experiment. The experimental results correspond well with the simulated results.

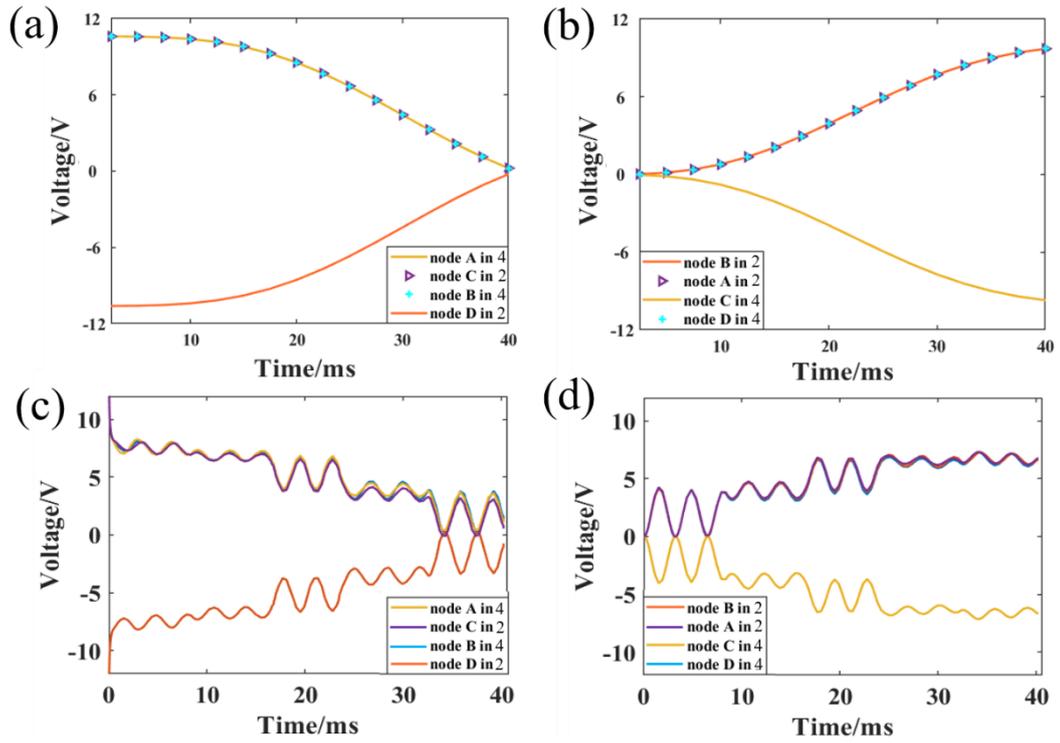

Figure S11 Voltages change with time in step IV to V. (a) and (b) are simulated results. (c) and (d) are experimental results.

**Supplementary Note 19. The braiding process of the two edges across a trivial segment**

Fig. S12 shows The braiding process of the two edges across a trivial segment.

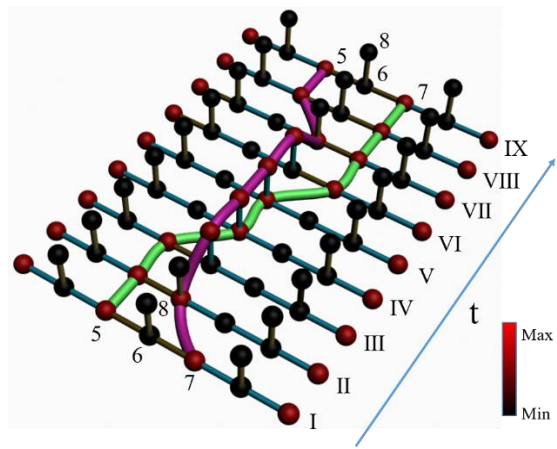

Figure S12 The braiding process of the two edges across a trivial segment.